\documentstyle[aas2pp4]{article}

\def \kms {km~s$^{-1}$}
\def \cubiccm {cm^{-3}}
\def \etal {et\ al.}

\def \solarmass {M_{\sun}}

\lefthead{Lee and Irwin} 
\righthead{HI in NGC~3044}

\begin{document}

\title{Neutral Hydrogen in the Edge-on Spiral Galaxy NGC 3044 -- Global
Properties and Discovery of HI Supershells}

\author{Siow-Wang Lee and Judith A. Irwin}
\affil{Department of Physics, Queen's University, Kingston, Canada K7L 3N6\\
	Email:swlee@astro.queensu.ca, irwin@astro.queensu.ca}

\begin{abstract}
The first detailed VLA mapping of the neutral hydrogen distribution in
the {\it isolated}, edge-on spiral galaxy NGC 3044 is presented.
Physical parameters such as $M_{HI}$, $M_T$, etc.  determined for this
galaxy are typical for galaxies of its morphological class (SBc).  We
have modelled the HI spectra in order to derive its global density and
velocity distributions.  An HI scale height of 420 $h^{-1}$ pc is thus
found.  This can be compared to the impressive radio continuum halo,
previously found to extend to $8~kpc$ above the midplane.

The present study reveals an asymmetry in the HI distribution as well
 as numerous high-latitude HI structures at various galactocentric
 radii.  The approaching (northwest) side of the galaxy is 14\% less
 massive than the receding side and its rotation curve does not reach
 terminal velocity. The rotation curve of the receding (southeast)
 side however resembles that of a normal galaxy.

Twelve high-latitude features were catalogued, of which four exhibit
the signature of an expanding shell.  There is some correlation of
these features with features observed in the radio continuum from
independent data.  The most massive shell (Feature 10) extends out to
6$h^{-1}$~kpc above the galactic disk. The radii and masses of these
shells range from $1.2~h^{-1} - 2.0~h^{-1}~kpc$ and $1.6\times10^7 -
5.5\times10^7h^{-2}~\solarmass$, respectively.  We have investigated
the possibility that the supershells could have been produced by
external impacting clouds, but conclude that this scenario is
unattractive, given the age of the shells, the isolation of the
galaxy, and the lack of any observed features sufficiently massive to
form the shells in the vicinity of the galaxy.  Therefore, an internal
origin is suggested.  Since the implied input energies from supernovae
are extremely high (e.g. from $1.4\times10^{53}h^{-2} -
7.4\times10^{55}h^{-2}~ergs$, corresponding to 400 $-$ 74,000
supernovae), we suggest that some additional energy (e.g.  from
magnetic fields) may be needed to produce the observed supershells.

\end{abstract}


\section{Introduction}
\label{sec:intro}

More than a decade ago, Heiles found HI shells, supershells and
``worms'' in the Milky Way Galaxy (Heiles 1979, 1984).  These
structures are seen protruding from the plane of the Galaxy in the
z-direction. Diameters of the shells and supershells range from a few
tens of pc to a few kpc. In some cases, the shells' diameters change
with velocity, indicative of expansion. Similar features are also
found in some external galaxies (e.g., NGC~5775, Irwin 1994; NGC~4631,
Rand and van der Hulst, 1993) either as shells, partial shells or
extensions. M31 is also observed to have hundreds of HI ``holes'',
regions devoid of neutral hydrogen, which are possibly cavities within
shells or supershells (\cite{bri86}). Several other example also
exist, but the list is still short.

Two possible explanations have been proposed: 1) the effect of stellar
winds and supernova explosions and 2) collisions of clouds with a
galactic disk (see \cite{ten88} and references therein).  Considerable
effort has been expended in the first category beginning with the
``galactic fountain'' model of Shapiro and Field (1976).  Norman and
Ikeuchi (1989) then extended this to develop the ``Chimney Model'' in
which they propose that high-latitude features are formed due to
clustered supernova explosions. The hot gas in the disk flows to the
halo via ``chimneys'' with walls of HI gas, cools at high-latitude and
subsequently falls back onto the disk. This way, a circulation of
mass, energy and momentum is set up between the disk and the halo. It
has been argued (\cite{ran93} and others) that kpc scale features
observed in our own and other galaxies are too large and require too
many input supernovae (e.g., $>$10,000 in some cases) to be explained
in this way. Thus there exists an energy problem for large HI
supershells if they are created by supernovae and stellar winds. The
cloud collision model, in part a response to this problem, on the
other hand, requires the presence of a companion. Thus the discovery
of HI shells in {\it isolated} systems would strongly favour an
internal origin. We have therefore targeted one such galaxy, NGC~3044,
which we felt to be a good candidate for HI extensions (see below) and
has no nearby companions.

\begin{figure}[ht]
\epsfxsize=2.3in \epsfbox[42 51 570 540]{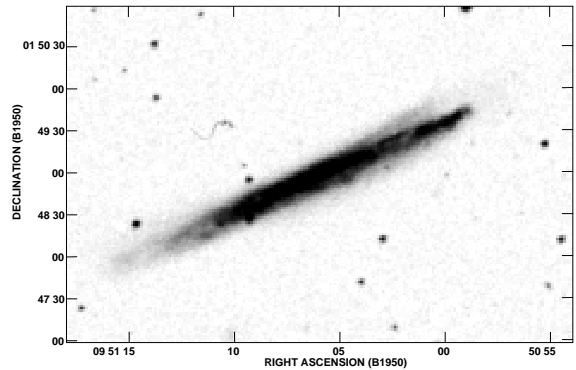}
\caption{Optical image of NGC~3044 obtained from the ST~ScI Digital
Sky Survey.} 
\label{f1}
\end{figure}

NGC~3044 (Fig.~\ref{f1}, from the ST~ScI Digital
Sky Survey \footnote{The Digitized Sky Surveys were produced
at the Space Telescope Science Institute under U.S.  Government grant
NAG W-2166. The images of these surveys are based on photographic data
obtained using the Oschin Schmidt Telescope on Palomar Mountain and
the UK Schmidt Telescope. The plates were processed into the present
compressed digital form with the permission of these institutions.}) 
is an edge-on, SBc galaxy in the Leo Cloud
(\cite{tul88}).  Solomon and Sage (1988) classify it as a Type I
galaxy (i.e, second lowest level of interaction), which has companions
within 10D$_{25}$ ($0.23h^{-1}$~Mpc in projection, where h is given by
$H_{\circ}$/100), on the Palomar Sky Survey (PSS) print, with a
velocity difference of less than 1000~\kms\ and showing no
morphological disturbance. A dust lane ($\alpha = 9^h51^m2\fs 0$,
$\delta = 1\arcdeg 49\arcmin 23\farcs 5$) appears prominently on the
northwestern (NW) side of the galaxy, obscuring part of the disk.
Above the dust lane, a faint feature protrudes at an angle away from
the major axis appears to be the source of the classification by
S\'{a}nchez-Saavedra $\etal$ (1990) of the galaxy as having a ``barely
perceptible'' optical warp pointing in a counter-clockwise direction
(inverse integral sign). On the southeastern (SE) side, the galaxy's
optical disk fades out more gradually than the NW end, suggesting a
possible asymmetric distribution of matter.

NGC~3044 was originally observed as part of a survey to search for
galaxies with extended radio continuum halos. The results
(\cite{sor94}) show the galaxy to have extended radio continuum
emission to distances as far as 8~kpc from the plane. Since this
galaxy is also infrared bright (\cite{soi87}), it therefore appeared
to be a good candidate for finding high-latitude neutral hydrogen arcs
and filaments. Throughout this paper, we use a distance to NGC~3044 of
16.14$h^{-1}$~Mpc (see \S\ref{subsec:glopro}). At this distance,
$1\arcsec = 0.078h^{-1}$~kpc.

In the following, we present the first high resolution HI observations
of NGC~3044. Thus, aside from searching for high latitude HI, we also
model the HI distribution in the disk, deriving global
parameters. Observations and data reduction are given in
\S\ref{sec:observations}, the HI distribution, velocity field and
the global HI profile of the galaxy in \S\ref{sec:results}.
\S\ref{sec:modelling} provides the details of modelling the galaxy
as well as a discussion of the modelling results. The observed
asymmetry and the energetics of the HI supershells are given in
\S\ref{sec:remarkable}. Finally, a summary can be found in
\S\ref{sec:summary}.

\section{Observations and Data Reduction}
\label{sec:observations}

NGC~3044 was observed with the Very Large Array
(VLA)\footnote{Operated by Associated Universities, Inc. under
contract with the National Science Foundation.} on June 19-20, 1993 in
the C configuration. The primary flux calibrators were 3C48 and 3C286,
having flux densities of 16.09~Jy and 14.88~Jy respectively. The
secondary calibrator, 0922+005 (flux density 0.75~Jy), was observed
about every half hour. The data were Hanning-smoothed
on-line. Bandpass calibration was done using the primary flux
calibrators. 33 line-free channels were averaged together and
subtracted from all 63 channels to obtain spectral line data alone.
The NRAO AIPS (Astronomical Image Processing Software) routine ``MX''
was used for mapping and deconvolution of the data cube (using the
``Cleaning'' algorithm). Two cubes were obtained, one using natural
weighting and the other using uniform weighting. Correction for
primary beam attenuation was applied to both cubes. Parameters
pertaining to the observations and mapping of NGC~3044 are listed in
Table~\ref{tab:observe}.

\begin{deluxetable}{lr}
\tablecolumns{2}
\tablewidth{0pt}
\tablecaption{HI Observation and Map Parameters \label{tab:observe}}
\tablehead{ \colhead{Parameter} & \colhead{Value}}
\startdata

VLA configuration & C \nl
Observing date & 1993 June \nl
On-source observing time & 8.75~hr  \nl
Approximate largest scale visible\tablenotemark{a} & 420$\arcsec$ \nl
Primary beam FWHM & 31$\farcm$5  \nl
Band center\tablenotemark{b} & 1318~\kms\ \nl
Total bandwidth & 1331~\kms\ \nl
Channel width (=resolution) & 20.8~\kms\ \nl
\sidehead{Synthesized beam parameters:}
Uniform weighting & 13$\farcs$6$\times$13$\farcs$5 @
PA=-78$\fdg$19  \nl
Natural weighting & 20$\farcs$7$\times$20$\farcs$2 @
PA=-29$\fdg$54  \nl
\sidehead{Root mean square map noise:}
Uniform weighting & 0.61~$mJy~beam^{-1}$ \nl
Natural weighting & 0.45~$mJy~beam^{-1}$ \nl
\sidehead{Rayleigh-Jean conversion factor ($T_b/S$):}
Uniform weighting & 3.30~$K(mJy~beam^{-1})^{-1}$ \nl
Natural weighting & 1.45~$K(mJy~beam^{-1})^{-1}$ \nl

\tablenotetext{a}{From Perley, 1994}
\tablenotetext{b}{Heliocentric, optical definition.}
\enddata
\end{deluxetable}

\section{Results}
\label{sec:results}

\subsection{The HI Distribution}
\label{subsec:HIdistri}

\subsubsection{The HI Channel Maps}

\begin{figure*}[ht]
\hspace{1in}\epsfxsize=4in \epsfbox[36 64 576 527]{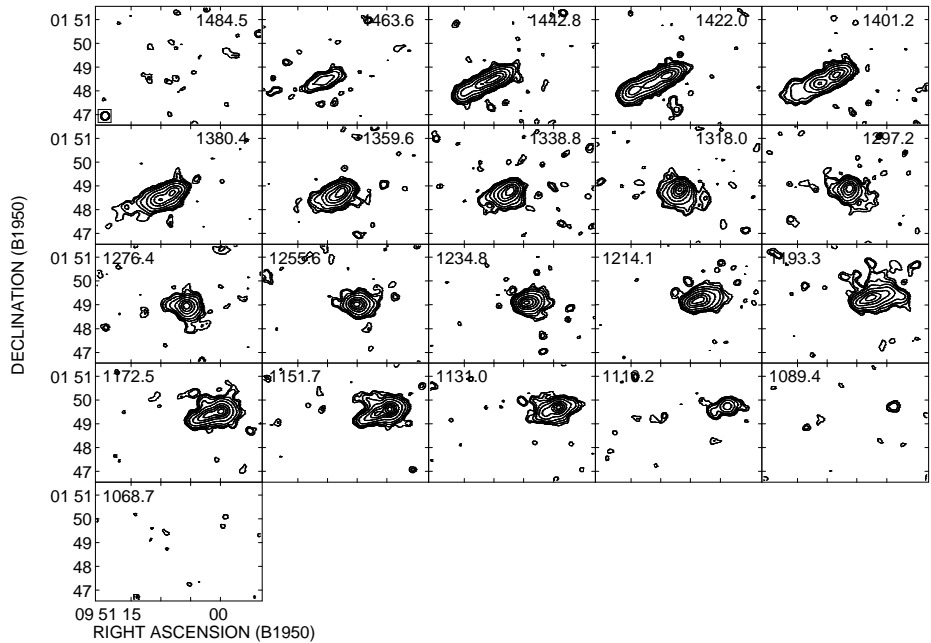}
\caption{(a) Naturally weighted velocity channel maps of
NGC~3044. Contour levels are at 1.0(1.5$\sigma$), 1.3, 1.9, 3.2, 6.4,
12.8, 19.2, 25.6, 32.0, 38.4, 44.8~mJy/beam. The channel width is
20~\kms\ centred at the velocity which appears at the upper left or
right corner of each frame. The synthesized beam is shown at the lower
left corner of the first frame.}
\label{f2a}
\end{figure*}

\addtocounter{figure}{-1}

\begin{figure*}
\hspace{1in}\epsfxsize=4in \epsfbox[36 165 576 610]{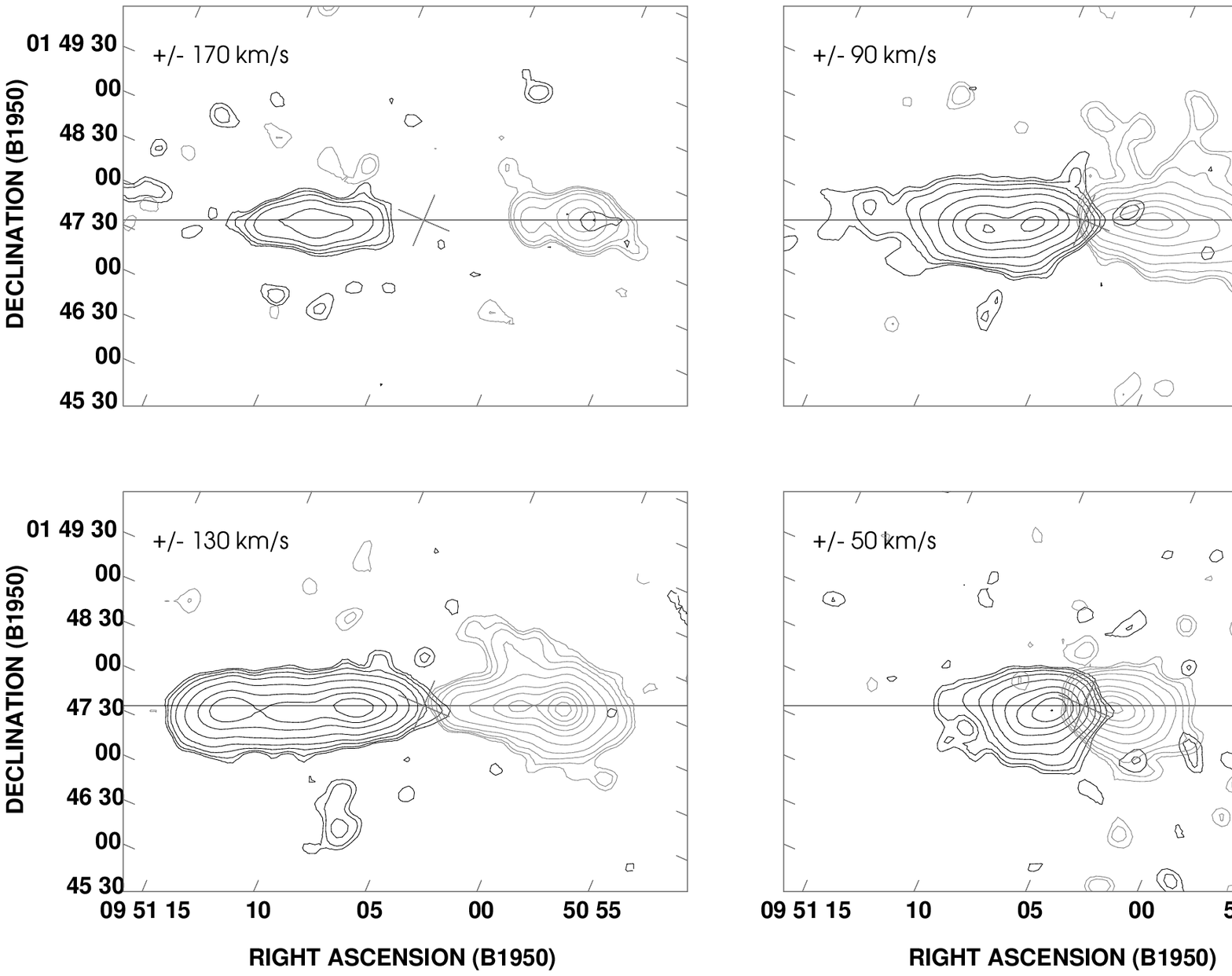}
\caption{(b) Selected velocity channels symmetrical with respect to
the systemic velocity are superimposed.  The rotation curve peaks at
150~\kms\ (Table~\ref{tab:model}). Dotted contours are positive in
velocity and solid contours are negative. The cross represents the
kinematic centre and the major axis is shown as the horizontal line
across. Contours are as in Fig. 2(a). Note that the images have been
rotated so that the major axis appears horizontal.}
\label{f2b}
\end{figure*}

The distribution of neutral Hydrogen in velocity space can most easily
be discerned from channel maps. The naturally weighted channel maps in
Figure~\ref{f2a}(a) show clearly that the SE side of the galaxy is
receding and the NW side is approaching. Since the dust lane occurs
most prominently along the south edge of the galaxy, this must be the
closest edge to us.  Therefore, any trailing spiral structure observed
should have an inverted ``s'' shape.

There is distinct evidence of HI arcs and extensions and high latitude
features away from the plane of the galaxy. For example, on the SE
side, a disconnected feature at $\alpha = 9^h51^m7\fs 7$, $\delta =
1\arcdeg 47\arcmin 21\farcs 0$ reaches a height of $8.4h^{-1}$~kpc in
the 1422~\kms\ channel and an impressive high-latitude extension is
found on the NW side at $\alpha = 9^h51^m3\fs 3$, $\delta =
1\arcdeg 50\arcmin 0\farcs 0$ from 1110~\kms\ to 1235~\kms. It
reaches a projected height of 7.1$h^{-1}$~kpc from the midplane. These
features and others will be discussed in more detail in
\S\ref{subsec:arcs}.

The channel maps also reveal the asymmetric distribution of gas in the
galaxy, which we highlight in Figure~\ref{f2b}(b), by superimposing
selected channel maps of equally red- and blue-shifted velocities with
respect to systemic. The HI distribution of the receding side (dotted
contours) is much more elongated than that of the approaching side
(solid contours) and the approaching side is more ``active''
displaying the largest high-latitude extensions above the HI disk. The
disk appears to be fairly straight at high velocities, although the
outer contours (1.5 - 3$\sigma$) on the NW side dip below the major
axis, probably due to an extension. The asymmetry will be discussed in
\S\ref{subsec:asymmetry}.

\subsubsection{The Column Density Maps}

\begin{figure*}[ht]
\hspace{1in}\epsfxsize=4in \epsfbox[36 100 576 573]{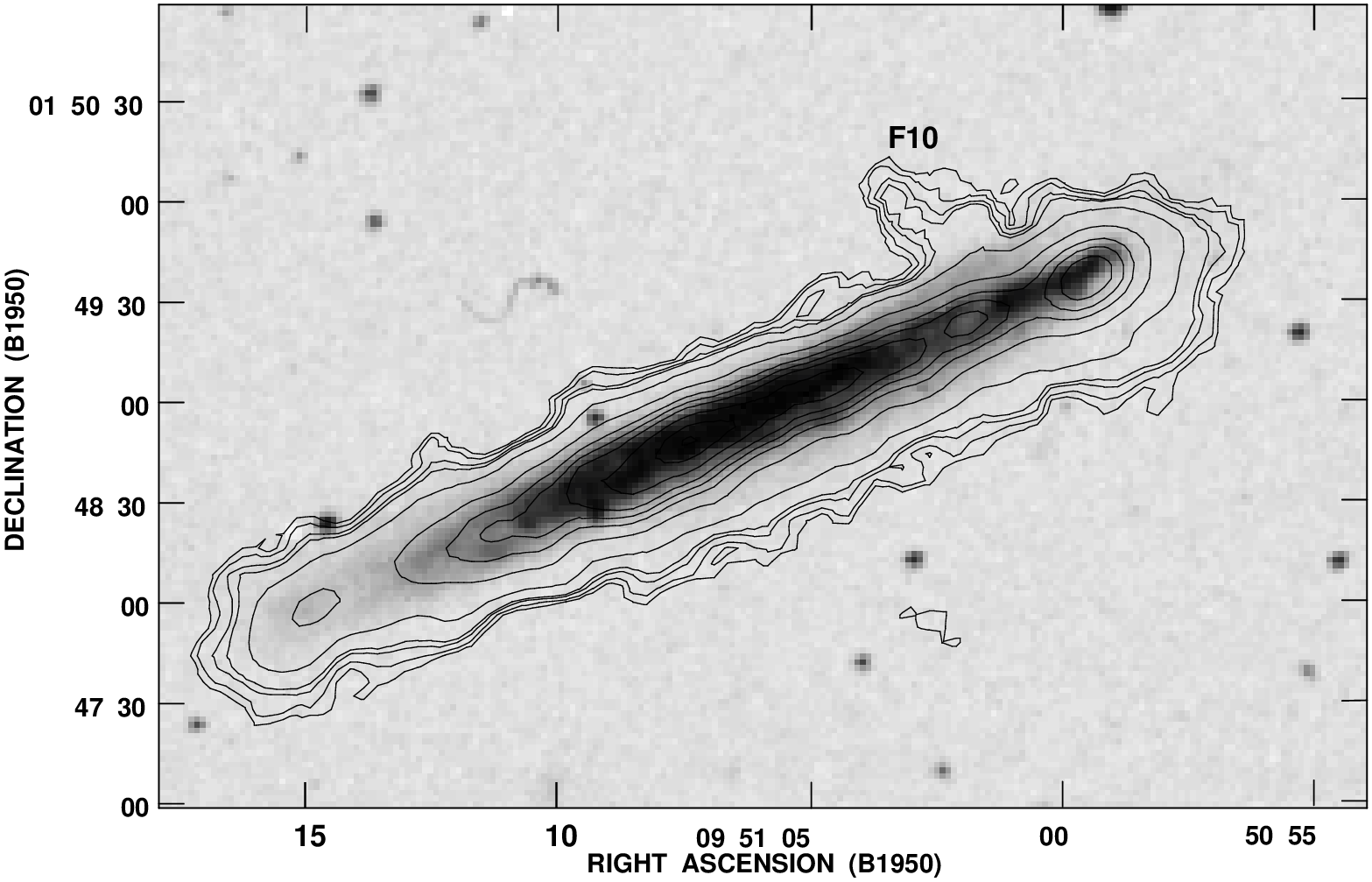}
\caption{(a) Uniformly weighted column density map superimposed on
the optical Digital Sky-Survey image (Fig.~\protect\ref{f1}). The map
is Hanning smoothed in velocity space using 3 channels and spatially
using a Gaussian of FWHM=20$\arcsec$. A cutoff at the 1$\sigma$ level
is applied. Contour levels are at 1.8, 3.0, 4.8, 6.0, 12.0, 30.1,
48.1, 60.0, 72.2, 90.2, 108.3,
120.3$\times$10$^{20}$~cm$^{-2}$. Feature 10 is labelled as F10.}
\label{f3a}
\end{figure*}

\addtocounter{figure}{-1}

\begin{figure*}[ht]
\hspace{1in}\epsfxsize=3in\epsfysize=3in \epsfbox[89 150 433 500]{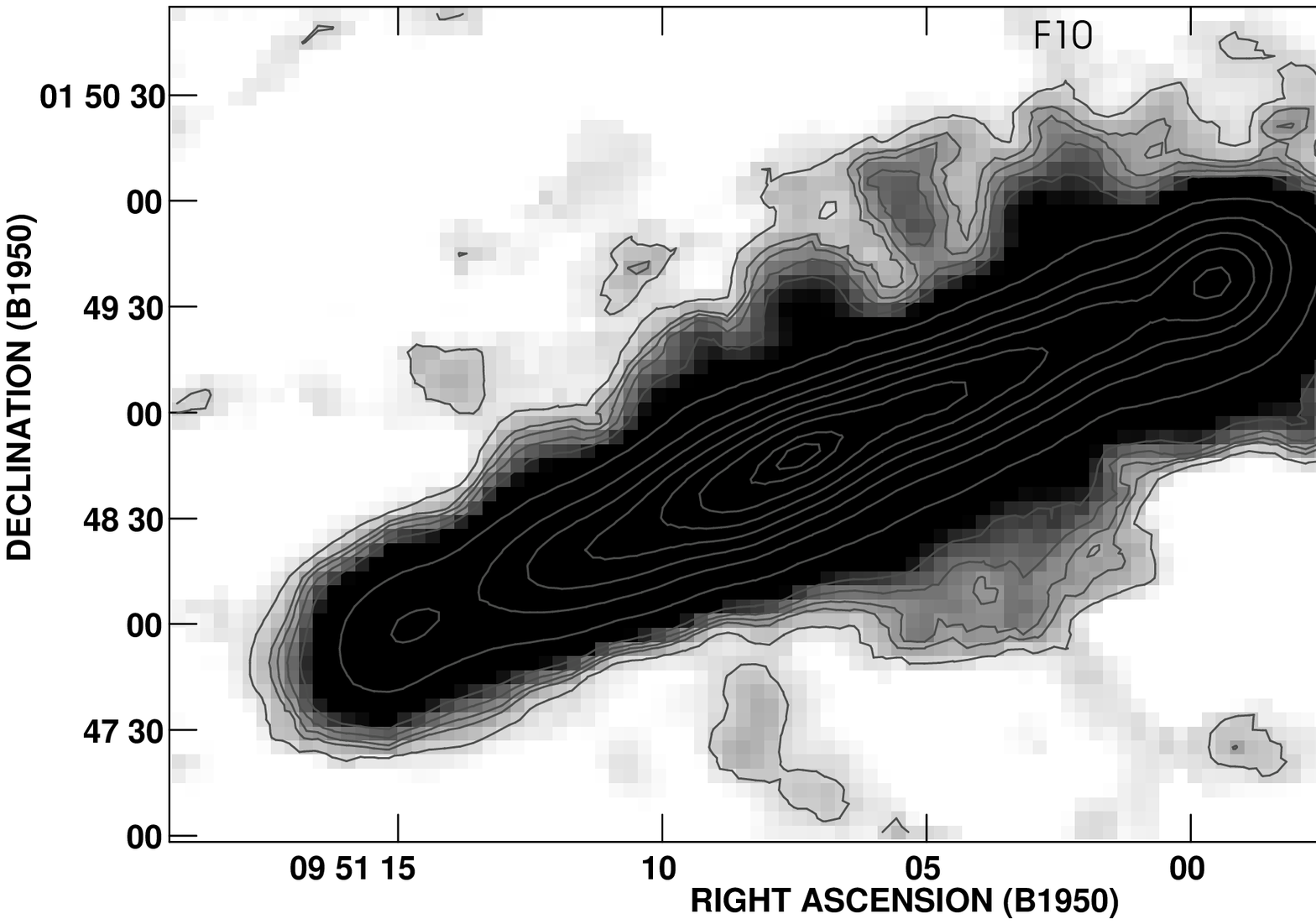}
\caption{(b) As in (a) but using naturally weighted cube. Column
density contours are superimposed on grey-scale map. Contour levels
are at 1.6, 2.6, 3.2, 4.0, 5.3, 13.2, 26.4, 39.7, 52.9, 66.1, 79.3,
92.5, 97.8\protect{$\times$10$^{20}$~cm$^{-2}$}.}
\label{f3b}
\end{figure*}

In Figure~\ref{f3a}(a), the velocity-integrated column density (moment
zero) map of NGC~3044 for the uniformly weighted data is superimposed
on the optical grey-scale image from the Digitized Sky Survey. The
naturally weighted column density map (contours and grey-scale) are
shown in Figure~\ref{f3b}(b). In Figure~\ref{f3a}(a), one extension is
particularly obvious -- labelled F10 (feature 10, see
\S\ref{subsec:arcs}). The disk's column density peaks at $\alpha =
9^h51^m7\fs 7$, $\delta = 1\arcdeg 48\arcmin 48\farcs 1$. Besides this
peak, there are three other peaks along the major axis. The two
outermost ones are roughly symmetrically located with respect to the
central peak, both at a distance of about 10$h^{-1}$~kpc
(126$\arcsec$) on either side. The NW peak is 2.5 times stronger in
intensity than its counterpart on the SE side, being $8.5\times
10^{21}~cm^{-2}$ and $3.3\times 10^{21}~cm^{-2}$, respectively.  The
peak of the HI distribution does not coincide with the peak of the
optical image but is offset by about 43$\arcsec$ along the major axis
to the SE. The optical distribution extends to the outer SE HI peak,
but with very low intensity.  Interestingly, there is another HI
column density enhancement (${\it{N}}_{HI} = 7.6\times
10^{21}~cm^{-2}$) between the central peak and the outer NW peak,
directly below feature 10. This smaller peak seems to coincide with a
region of low surface brightness on the optical image, which could be
due to absorption by a dust lane.

Measured from the outermost contour of the naturally weighted map
[Fig.~\ref{f3b}(b)], the HI disk spans a length of 5$\farcm$7 or
2.3$\times$R$_{25}$ along the major axis. Feature 10 is clearly
visible and more extended than in Figure~\ref{f3a}(a) along with
various other extensions. F10 extends out to about 6$h^{-1}$~kpc above
the plane of the galaxy in Figure~\ref{f3b}(b). In fact,
Figure~\ref{f3b}(b) shows an extensive disturbance in the region near
feature 10.  We show below (see $\S~\ref{subsec:arcs}$) that feature
10 is indeed an expanding feature. Other protruding features are also
visible in Figure~\ref{f3b}(b). For example, the hole at $\alpha =
9^h51^m4\fs 0$, $\delta = 1\arcdeg 48\arcmin 10\arcsec$ and numerous
protrusions along the northern edge of the galaxy can be seen.

\subsection{The Velocity Field}
\label{subsec:velocity}

\begin{figure}[ht]
\epsfxsize=2.3in \epsfbox[42 126 569 656]{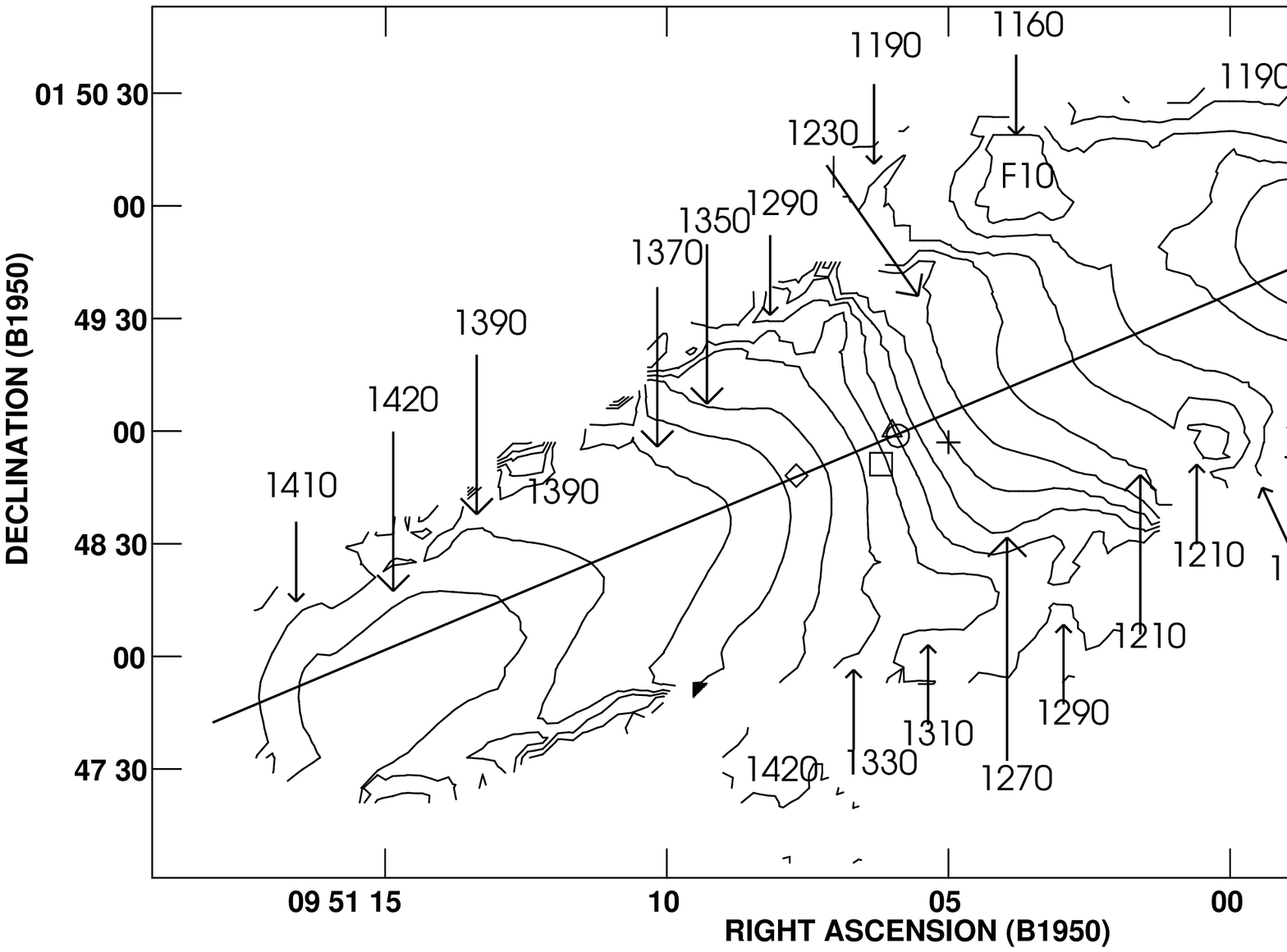}
\caption{The naturally weighted velocity field of NGC~3044.
Contours are 20~\kms\ apart starting from 1130~\kms\ to 1410~\kms.
Also included are contours of 1160 and 1420~\kms. The kinematic
major axis is indicated by the straight line across. Symbols represent
the different peaks and centres described in the text. + is the
optical centre, $\triangle$ is the radio continuum peak, $\diamond$ is
the column density peak, $\bigcirc$ is the centre based on the global
HI profile and $\Box$ is the model centre.}
\label{f4}
\end{figure}

The naturally weighted velocity field of NGC~3044 is shown in
Figure~\ref{f4}. The kinematic major axis is parallel to the optical
major axis to within one degree out to a radius of about 50$\arcsec$
(4$h^{-1}$~kpc). The kinematic minor axis is not perpendicular to the
inner kinematic major axis, this is typically an indication of a bar
structure (Bosma, 1981), which is consistent with the optical
classification of this galaxy. Beyond about 1$\farcm$7 (8$h^{-1}$~kpc)
on the NW side, the velocity field becomes disturbed. For example,
there is a closed 1160~\kms\ contour at the position of F10. On both
sides, beyond 4$h^{-1}$~kpc, the kinematic major axis bends slightly
towards the south.  Therefore the major axis does not resemble the
more commonly seen kinematic warp in other edge-on galaxies and we
interpret the disturbance near feature 10 as high-latitude features
instead of a simple warp.

In Figure~4, we plot the systemic velocity of the galaxy, obtained
from the global profile, on the major axis of the galaxy (represented
by $\bigcirc$ at $\alpha = 9^h51^m5\fs 9$, $\delta = 1\arcdeg
48\arcmin 58\farcs 7$). This position is significantly different from
the position of both the column density peak (labelled $\diamond$ in
Fig.~4) as well as the optical centre (labelled +), so that the
kinematic centre is between the column density peak and optical
centre. However, the kinematic centre does coincide with the peak of
the radio continuum map (represented by $\triangle$).  The kinematic
centre given by our model (labelled $\Box$) is also shown. This will
be discussed more fully in \S\ref{subsec:modelgeo}.  Here, we
identify the radio continuum centre as the `true' centre of the
galaxy.

Notice all velocity contours have bends and kinks in them, especially
near the outer edges of the contours. Such features are usually
explained by shock fronts occurring at the inside edge of spiral arms
as a density-wave perturbs the local velocity field
(\cite{vis80}). However, due to the high inclination of NGC~3044, this
is difficult to verify from the column density map. At the location of
feature 10, the velocity field is inconsistent with normal galactic
rotation; instead it is more blueshifted.  If feature 10 is actually
located at a larger distance (so that its distance from the nucleus in
Figure~\ref{f4} is just a foreshortened distance in projection), then
we would expect its radial velocity to be redshifted. It is clear that
feature 10 does not follow the general flow of material in the disk
below it.

\subsection{The HI Global Profile}
\label{subsec:glopro}

\begin{figure}[ht]
\epsfxsize=3in \epsfbox[20 17 592 779]{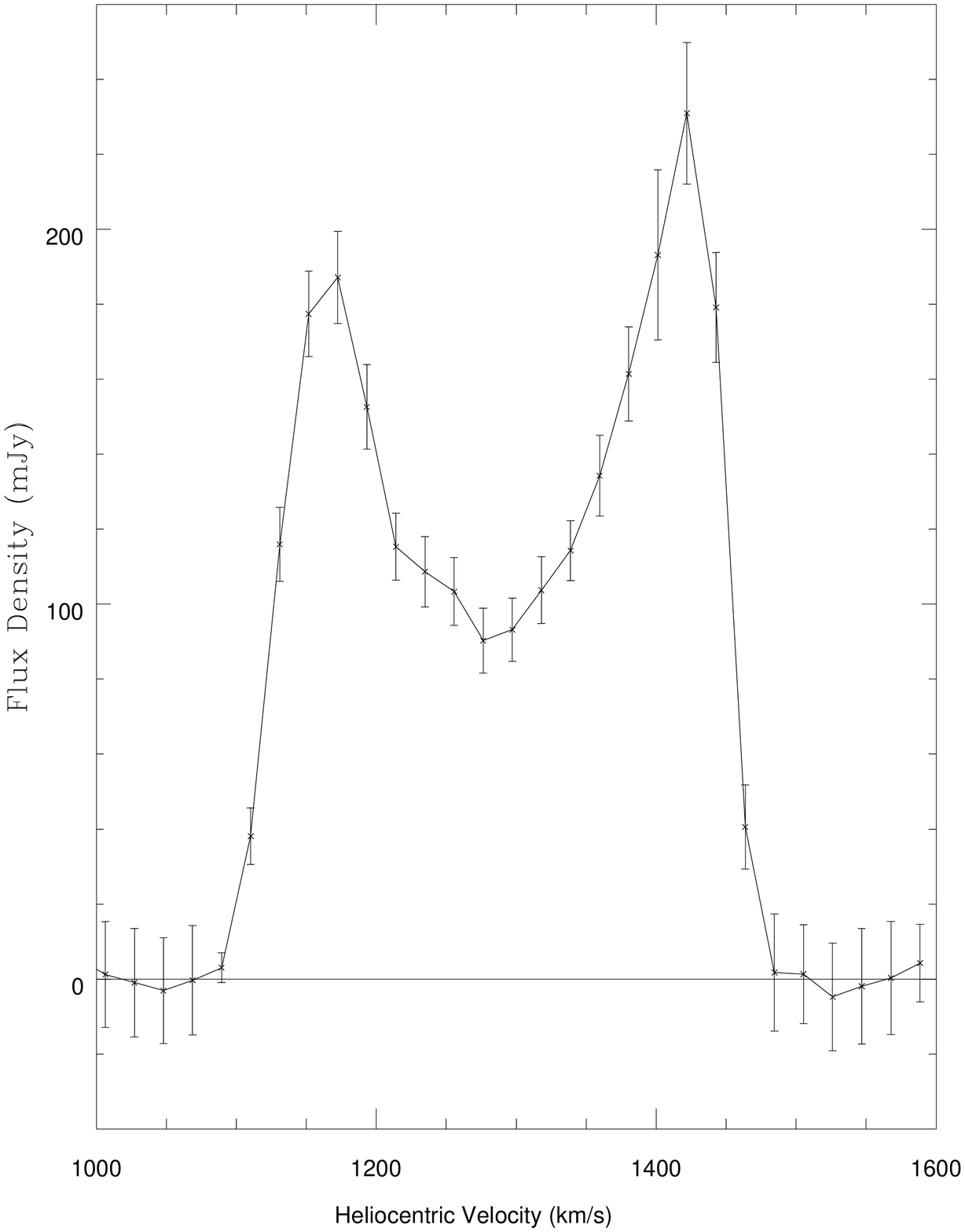}
\caption{The naturally weighted global profile for NGC~3044.
Error bars are at the 1$\sigma$ level.}
\label{f5}
\end{figure}

Figure~\ref{f5} shows the global profile of NGC~3044 from the
naturally weighted cube. The profile agrees within error with
previously published data (\cite{sta88}, 12$\arcmin$ beam).

The global properties of NGC~3044 are tabulated in
Table~\ref{tab:glopara}. Row~1 gives the systemic velocity from the
midpoint of the 20\% intensity level. Row~2 gives distance deduced
from the systemic velocity corrected to the reference frame of the 3K
background radiation (see the Third Reference Catalogue of Bright
Galaxies [\cite{deV91}], hereafter RC3). Row~3 gives the velocity
width at the 20\% level uncorrected for inclination due to the high
inclination for this galaxy ($i = 84\arcdeg \pm 2\arcdeg$,
\cite{bot84}).

\begin{deluxetable}{clc}
\tablecolumns{3}
\tablewidth{4in}
\tablecaption{Global HI Properties of NGC~3044 \label{tab:glopara}}
\tablehead{\colhead{Row} & \colhead{Parameter} & \colhead{Value}}
\startdata

(1) & $V_{sys}$ (\kms) & 1287 $\pm$ 10 \nl 
(2) & $D_{Hubble}$ ($h^{-1}$~Mpc) & 16.14 \nl 
(3) & $\Delta V_{20\%}$ (\kms) & 351 $\pm$ 10 \nl 
(4) & $\int{S\cdot dV}$ (Jy$\cdot$kms) & 48.2 $\pm$ 3.6 \nl
(5) & $M_{HI} (h^{-2}~10^9~M_{\sun})$ & 3.0 $\pm$ 0.2 \nl
(6) & $M_{HI}/L_B (M_{\sun}/L_{\sun})$\tablenotemark{a} & 
		0.22 $\pm$ 0.02 \nl 
(7) & $M_T (h^{-1}10^{11}~M_{\sun})$ \tablenotemark{b} & 
	1.1 $\pm$ 0.1 \nl
(8) & $M_T/L_B (h~M_{\sun}/L_{\sun})$ & 8.1 $\pm$ 0.4 \nl 
(9) & $M_{HI}/M_T (h^{-1})$  & 0.027 $\pm$ 0.002 \nl 
\enddata
\tablenotetext{a}{$\it{L}_B$ is calculated using the total ``face-on''
apparent blue magnitude ($\it{B}^T_o$) given in the RC3 and using a
value of +5.48 for the absolute blue magnitude of the Sun.}
\tablenotetext{b}{Radius is taken to be the maximum extent of the
third contour in the naturally weighted position-velocity diagram
(Fig.~\protect\ref{f6}) and $\it{V}_{rot}$ is the average of the
maximum blue- and red-shifted velocities with respect to the systemic
velocity measured using the same contour. The third (4.5$\sigma$)
contour is used so as to avoid the protruding feature on the
approaching side of the p-v diagram (see Fig.~\ref{f6}) which
probably does not reflect the maximum blueshifted rotational
velocity.}
\end{deluxetable}

The integrated flux density of NGC~3044 is given in row~4. Our value
agrees with both Staveley-Smith and Davies (1988) and Krumm and
Salpeter (1980). Row~5 shows the estimated HI mass of the galaxy using
the equation $\it{M}_{HI} = 2.35\times 10^5 D^2 \int{S\cdot dV}$ where
$\it{M}_{HI}$ is in $\it{M}_{\sun}$, $D$ is in Mpc and $\int {S\cdot
dV}$ is in Jy$\cdot$\kms. This equation is valid under the assumption
of optical thinness. Haynes and Giovanelli (1984) investigated the
effect of HI self-absorption on HI integrated flux as a function of
morphological type and galaxy inclination. They concluded that for
Sc-type galaxies with inclinations like NGC~3044, the correction
factor is about 1.3. Thus $\it{M}_{HI}$ could be up to 30\% higher
than the value in Table~\ref{tab:glopara}.  The peak brightness
temperature for NGC~3044 is 83.1~K in the 13$\arcsec$ beam. Thus
optical depth effects should not be strong.  The HI mass of the
receding side is 14\% larger than the approaching side while the
uncertainty due to the error bars in the global profile is only
8\%. This asymmetry is obvious in Fig.~\ref{f5} and will be discussed
in \S\ref{subsec:asymmetry}.

Row~6 in Table~\ref{tab:glopara} gives the HI mass to blue luminosity
ratio.  Row~7 gives the estimated total mass of the galaxy within the
outermost detectable HI radius. The usual equation assuming a
spherical geometry is used [$\it{M}_T = 2.33\times
10^5\it{R}\it{V}_{rot}^2 (\it{M}_{\sun})$].  The quoted uncertainty
reflects only the asymmetry of the receding and the approaching sides
of the galaxy, which dominates the errors. Using a spherical geometry
may result in an overestimate of the true total mass within $\it{R}$
if the total mass is dominated by the disk component. The overestimate
will be no more than 40\% (Lequeux 1983).  Row~8 gives the ratio of
total mass to blue luminosity and row~9 gives the fractional HI mass.

The values of Table~\ref{tab:glopara} were compared with those of Roberts and Haynes
(1994) for the Local Supercluster (galaxies with $v<3000$~\kms),
correcting to their value of $H_{\circ}$. With an additional slight
correction for the method of calculating $M_T$, we find that all
values for NGC~3044 are typical for its Hubble type.

\section{Data Cube Modelling and Results}
\label{sec:modelling}

For edge-on galaxies, the usual first and second moment analysis
(\cite{van94}) cannot be used successfully to obtain density and
velocity distributions because the flux density at each pixel results
from an integration along the line of sight which spans many
galactocentric radii.

We instead use the technique of Irwin and Seaquist (1991) and Irwin
(1994) which models the intensity at every pixel given the radial and
perpendicular density distributions and the form of the rotation
curve. These curves are parameterized, and the parameters are fit via
a non-linear least squares algorithm. The parameters are a subset of
{\it{RA}} and {\it DEC} of the nucleus, the position angle ({\it PA})
and inclination ({\it i}) of the galaxy, the systemic velocity
({$\it{{V}_{sys}}$}), the galactocentric radius ({${\it{R}_{max}}$})
at which maximum rotational velocity ({$\it{{V}_{max}}$}) occurs, a
rotation curve shape index ({\it m}), the peak volume density of the
distribution ({$\it{n_{max}}$}), the density scale length for the
radial distribution ({$\it{r_o}$}) and the scale height ({$\it{z_o}$})
of the vertical distribution. The model intensities are then smoothed
to the spatial resolution of the data and the residuals (data minus
model) are found. The best result is considered to be the model
producing the lowest residuals.

\begin{figure}[ht]
\epsfxsize=3in \epsfbox[36 92 576 699]{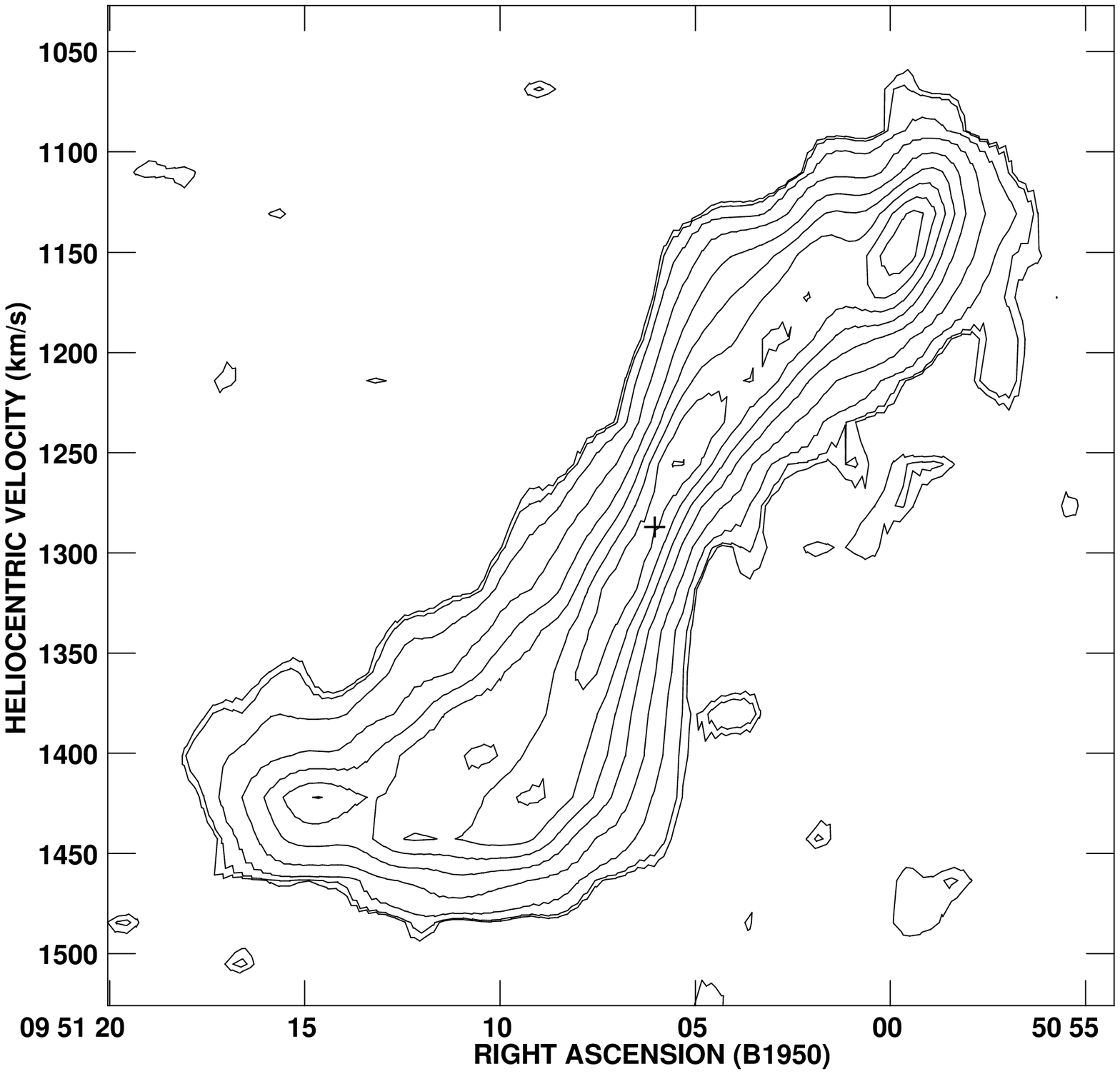}
\caption{Position-velocity diagram along the kinematic major
axis of NGC~3044. The data cube used is naturally weighted. Contour
levels are 0.8(1.8$\sigma$), 1, 2, 5, 10, 15, 20, 25, 30, 35,
40~mJy/beam.  The cross indicates the radio continuum centre placed at
$V_{sys}$ based on the global HI profile. Note that at the galaxy's 
declination and P.A., 1$^s$ corresponds to 16$\farcs$3 along the 
major-axis. Same for all similar figures hereafter.}
\label{f6}
\end{figure}

The trial density distributions are exponential or Gaussian for both
the radial and z-axis distributions centered at the galaxy centre and
midplane, respectively. In addition, a Gaussian or exponential ring
distribution in the plane of the galaxy centered at a radius of
{$\it{R}_o$} was also tried. The choices for the form of the velocity
curve are the Brandt curve and a user specified numerical rotation
curve which we take from the velocity-position diagram along the major
axis (Fig~6) (allowing the amplitude to vary). The model also allows
for a velocity dispersion. Thus the geometrical parameters
(particularly the inclination which cannot normally be found without
an assumption of the scale height), the velocity distribution and the
density distribution (radial and vertical) can be found. For NGC~3044,
we model the uniformly weighted data cube as it has a higher spatial
resolution than the naturally weighted cube, yet the integrated flux
density (42$\pm$6~Jy$\cdot$km/s) still agrees with that of the
naturally weighted cube (see Table~\ref{tab:glopara}). There are 2248
independent data points of real positive emission, hence the free
parameters should be well constrained.

The best fit results are shown in Table~\ref{tab:model} for the entire
galaxy (column~2) and the receding (column~3) and advancing sides
(column~4). Column~1 is a list of free parameters. Brandt rotation
curves are used for all the models in this table, since they resulted
in better fits in all trials than a user-specified numerical curve. We
shall hereafter use notations such as `RG(GS)-EX' to represent a model
having a Gaussian ring distribution in the plane and an exponential
distribution in the z-direction. The uncertainties quoted are either
standard deviations of the results of all trial models (including
those not shown in Table~\ref{tab:model}) or formal 1$\sigma$ errors
from the fit of the best model, whichever is larger. The following
subsections discuss the model results.

\begin{deluxetable}{lrrr}
\tablewidth{0pt}
\tablecaption{Model Parameters \label{tab:model}}
\tablehead{
\colhead{Parameter} & \colhead{Whole Galaxy} & \colhead{Receding Side} &
\colhead{Approaching Side} \nl
\colhead{} & \colhead{RG(GS)-GS} & \colhead{GS-GS} & \colhead{RG(GS)-EX} \nl
\colhead{} & \colhead{ring @ 30$\arcsec$} & \colhead{} & \colhead{ring @
104$\arcsec$} \nl
\colhead{(1)} & \colhead{(2)} & \colhead{(3)} & \colhead{(4)}}
\startdata
\nl
$\alpha_o$(1950) & $9^h51^m6\fs 15\pm0\fs07$ & \nodata & \nodata \nl
$\delta_o$(1950) & $1\arcdeg48\arcmin 51\farcs 2\pm0\farcs4$ & \nodata & \nodata \nl
PA($\arcdeg$) & 112.95$\pm$0.10 & 113.45$\pm$0.04 & 111.4$\pm$0.3 \nl
$i(\arcdeg)$ & 84.9$\pm$0.4 & 85.3$\pm$0.3 & 86.1$\pm$0.2\nl
$V_{sys}$(\kms) & 1298$\pm$4 & \nodata & \nodata \nl
$V_{max}$(\kms) & 150$\pm$3 & 149$\pm$2 & 151$\pm$3 \nl
$R_{max}(h^{-1}~kpc)$ & 6.4$\pm$0.4 & 4.9$\pm$0.2 & 9$\pm$4 \nl
$R_o(h^{-1}~kpc)$ & 2.3 & \nodata & 8.2 \nl
m & 0.99$\pm$0.18 & 0.64$\pm$0.15 & 0.48$\pm$0.26 \nl
$n_{max}(h~\cubiccm)$ & 0.38$\pm$0.15 & 0.40$\pm$0.18 & 0.33$\pm$0.16 \nl
$r_o(h^{-1}~kpc)$ & 5.3$\pm$0.7 & 6.4$\pm$0.8 & 1.9$\pm$0.2 \nl
$z_o(h^{-1}~kpc)$ & 0.42$\pm$0.04 & 0.49$\pm$0.04 & 0.45$\pm$0.03 \nl
$r_{o,i}(h^{-1}~kpc)\tablenotemark{a}$ & 1.20$\pm$0.02 & \nodata & 
2700$\pm$29 \nl
$\sigma_v$ (\kms) & 24.3$\pm$2.2 & 19.9$\pm$2.2 & 17.7$\pm$2.2 \nl
\nl

\enddata
\tablenotetext{a}{Inner Scale Length for $r<R_o$.}
\end{deluxetable}

\subsection{Galaxy's Geometry}
\label{subsec:modelgeo}

The position of the nucleus of the galaxy is well constrained amongst
the different models giving a standard deviation of 0$\fs$07 in RA and
0$\farcs$4 in DEC. This is not surprising since the kinematic centre
of a galaxy does not depend on the form of the density distributions
either parallel or perpendicular to the plane. As mentioned before,
the column density peak is shifted towards the SE (receding side)
(\S\ref{subsec:velocity}). As part of the modelling process, we
forced the model's nucleus to coincide with the column density peak
and the radio continuum peak separately. These resulted in worse
models for all combinations of density distributions implying real
offsets of these peaks from the model's centre. The kinematic centre
as given by the best model is 2$\farcs$3 east and 8$\farcs$8 south of
the radio continuum peak (9$^h$51$^m$6$\fs$0,
1$\arcdeg$49$\arcmin$0$\farcs$0, see Fig.~\ref{f4}).  As the receding
side is more massive, it is natural that the model's centre is being
weighted towards this side, hence the offset of the model's centre
from the radio continuum centre. The peak of the column density map is
23$\farcs$3 east and 3$\farcs$1 south of the modelled galaxy centre.
These differences in the peak positions are significant. The P-V
diagram of the major axis of the galaxy (Fig.~\ref{f6}) also shows an
asymmetry in the sense that the receding side of the galaxy extends
further. In addition, the flat part of the rotation curve shows up
more prominently on this side.  It therefore appears that either the
galaxy's HI emission has been `truncated' on the approaching side or
the receding side emission has been `stretched out'. We feel that the
former is probably true as the receding side of the P-V diagram more
closely resembles the P-V diagrams of other giant spirals.

The position angle and inclination of the galaxy are also well
determined.  Various models resulted in standard deviations of
0$\fdg$10 and 0$\fdg$4 for these two parameters, respectively. The
position angle agrees with that given in the RC3 (note that there is a
mistake in the RC3, the position angle listed should be 113$\arcdeg$
instead of 13$\arcdeg$). The inclination from the model ($84\fdg 9$)
agrees with Bottinelli $\etal$ (1984) and Staveley-Smith and Davies
(1988) who obtain {\it{i}}=84$\pm$2$\arcdeg$ and 90$\pm$13$\arcdeg$
respectively. The former assumes an intrinsic oblateness of 0.15 while
the latter uses 0.20. Our value is independent of the intrinsic axial
ratio, and should be an improvement over the previous estimates.

\subsection{The Rotation Curve}

The systemic velocity given by the best model agrees with that
measured from the global profile of the data (Table~\ref{tab:glopara})
to within errors. Fixing the systemic velocity at a lower value
resulted in worse models for all trials.  The rotation curve
parameters, {\it{V$_{max}$}} and {\it{R$_{max}$}}, are also
well-defined, with the spread in {\it{V$_{max}$}} within 10\% of the
velocity resolution while the spread in {\it{R$_{max}$}} is determined
to 40\% of the HPBW. The shape of the rotation curve is determined by
the Brandt index, {\it{m}}, and the spread in this parameter is only
18\% amongst all trials. When a velocity dispersion is included in the
models, results for every model improved significantly. For the best
model, the optimum FWHM of the Gaussian smoothing function is found to
be 57.2~\kms\ (2.75 channels) corresponding to a velocity dispersion of
24.3~\kms. This value reflects all non-circular motions along each
line of sight, including any velocity variations due to spiral arms
and is a global average. This velocity dispersion is comparable to
that found in NGC~5775 (Irwin, 1994) and is about 3 times the velocity
dispersion found for the Milky Way galaxy, which only measures the rms
cloud velocity (Spitzer, 1978).

\subsection{The Density Distributions}
\label{subsec:modeldensity}

The model parameters related to the density distributions vary
significantly amongst different models. This is expected as the
parameters are defined differently between some of the models. While
there is no obvious central hole in the column density map (Figs.~3),
the best results of the modelling are found using a ring distribution
rather than a distribution which peaks at the centre. From our best
model (Gaussian ring in-plane and Gaussian in z-direction), the volume
density at the radius of the ring (2.3$h^{-1}$~kpc) is found to be
0.38$h$~$\cubiccm$. This is very similar to the average midplane
density of our Galaxy between 4 to 8~kpc which is
$\approx$0.35~$\cubiccm$ (\cite{bur78}). The outer radial scale length
is 5.3$h^{-1}$~kpc, or 20\% of the length of the HI disk as measured
by the outermost contour of the column density map
[Fig.~\ref{f3b}(b)]. As a comparison, the Galactic HI density stays
roughly flat from about 4 to 10~kpc and falls off beyond that. The
inner scale length is the least constrained, its value depends
sensitively on the position of the ring. The uncertainty associated
with this parameter in Table~\ref{tab:model} is therefore taken to be
the 1-$\sigma$ uncertainty of the fit instead of the standard
deviation amongst the different models. Our best model gives the inner
scale length of 1.20$h^{-1}$~kpc, about half the radius of the ring.

\begin{deluxetable}{lccccc}
\tablewidth{0pt}
\tablecaption{High-Latitude Arcs and Extensions \label{tab:shells}}
\tablehead{
\colhead{Feature} & \colhead{Right Ascension} & \colhead{Declination} &
\colhead{Velocities} & \colhead{Number of} & \colhead{Height from midplane} \nl
\colhead{} & \colhead{$^h~~^m~~^s$} & \colhead{$\arcdeg~~ \arcmin~~ \arcsec$} &
\colhead{\kms} & \colhead{Channels} & \colhead{$h^{-1}$~kpc}}
\startdata
\nl
1 & 9 51 9.8 & 1 49 5.0 & 1401.2 to 1442.8 & 3 & 2.9 \nl
2 & 9 51 6.9 & 1 49 21.0 & 1359.6 to 1422.0 & 4 & 3.5 \nl
3 & 9 51 5.0 & 1 48 17.0 & 1234.8 to 1359.6 & 7 & 6.0 \nl
4 & 9 51 10.6 & 1 48 13.0 & 1338.8 to 1297.2 & 3 & - \nl
5 & 9 51 7.7 & 1 47 21.0 & 1380.4 to 1463.6 & 5 & 8.4 \nl
6 & 9 51 9.5 & 1 49 5.0 & 1255.6 to 1318.0 & 4 & 5.0 \nl
7 & 9 51 4.7 & 1 48 17.0 & 1276.4 to 1318.0 & 3 & - \nl
8 & 9 51 6.9 & 1 49 29.0 & 1193.3 to 1276.4 & 5 & 3.7 \nl
9 & 9 51 5.3 & 1 50 1.0 & 1193.3 to 1214.1 & 2 & 6.6 \nl
10 & 9 51 2.6 & 1 50 0.0 & 1110.2 to 1234.8 & 7 & 7.1\nl
11 & 9 50 58.9 & 1 50 2.0 & 1151.7 to 1214.1 & 4 & 4.9 \nl
12 & 9 50 57.8 & 1 49 0.0 & 1131.0 to 1193.3 & 4 & 5.2 \nl
\nl

\enddata
\end{deluxetable}

The vertical scale height given by the best model is $0.42h^{-1}$~kpc.
This value is obtained as a global parameter, an average over the
entire disk of the galaxy. Note that in specific places, e.g., feature
10, the HI extends to 5.7$h^{-1}$~kpc in projection. In the Galaxy, it
is known that the HI distribution between 4 to 8~kpc consists of two
components, a central layer with a FWHM of $\approx$ 0.1~kpc, and a
low-intensity, high-temperature component with a FWHM of
$\approx$0.5~kpc (Lockman, 1984). It is perhaps more helpful to
compare the vertical scale height of NGC~3044 with an external
galaxy. NGC~891 is also an edge-on, IR-bright spiral which exhibits
high-latitude HI features. van der Kruit (1981) modelled the thickness
of its HI layer and found that the FWHM of the z-distribution
increases with galactocentric radii. The FWHM varies from 0.32~kpc at
a radius of 4.2~kpc to 1.89~kpc at a radius of 20.8~kpc with an
average value of 0.9~kpc. Therefore, NGC~3044 appears to have a
moderately thick HI disk compare to the Galaxy yet not as extensive as
that in NGC~891. To check the validity of a thick disk, we modelled
the galaxy again by fixing the vertical scale height to a low value
(0.1~kpc).  This resulted in a significantly worse model fit to the
data. In addition, Sorathia (1994) finds a large radio continuum scale
height of 1.8$\pm$0.5~kpc for the 20~cm C-array data (see
Fig.~9). Thus NGC~3044 has both a moderately thick HI disk as well as
a thick radio continuum disk.

\subsection{Modelling of Receding and Approaching Halves of the Galaxy}

As we have noted before, the HI distribution of NGC~3044 is not
symmetrical in the sense that the HI is more extended on the SE
(receding side) and less so on the NW (approaching side). Modelling
the receding and approaching side of the galaxy separately enforces
this picture. Table~\ref{tab:model} lists the best fit parameters for
the two halves of the galaxy in comparison with the parameters given
by the galaxy as a whole.  Column 3 and 4 were obtained by fixing the
nucleus position and the systemic velocity given by the best model for
the whole galaxy (Table~\ref{tab:model}, column 2).

The model fits the receding side of the galaxy better which shows that
this side of the galaxy is better described by a smooth spatial and
velocity distribution of gas. This result simply reiterates what we
see in the major axis rotation curve from the data (Fig.~\ref{f6}).
Although $\it{{V}_{max}}$ and $\it{{R}_{max}}$ of both sides agree
within error, the uncertainty in $\it{{R}_{max}}$ is much larger for
the approaching side. The Brandt curve indexes show the same
trend. The difficulty in pinning down these parameters is due to the
fact that on the approaching side, the rotation curve never reaches a
terminal velocity. For the density distribution parameters, the two
halves of the galaxy differ significantly.  The receding side is best
fitted by a GS-GS distribution while the approaching side prefers a
RG(GS)-EX distribution. In the latter case, the location of the ring
is at the outer density peak and the inner scale length is extremely
high, hence the HI distribution inside this peak is essentially
flat. A consequence of the large ring radius is the small outer scale
length, as expected. These results for both sides, separately, are
consistent with the residual emission observed in Fig.~\ref{f7} (see
next section).  Finally, the approaching side has a smaller velocity
dispersion than the receding side, which can also be seen from the
narrower (in velocity) emission on the approaching side in Fig.~6.

\subsection{Comparing Model With Data}
\label{subsec:residual}

\begin{figure}[ht]
\epsfxsize=2in \epsfbox[100 262 501 550]{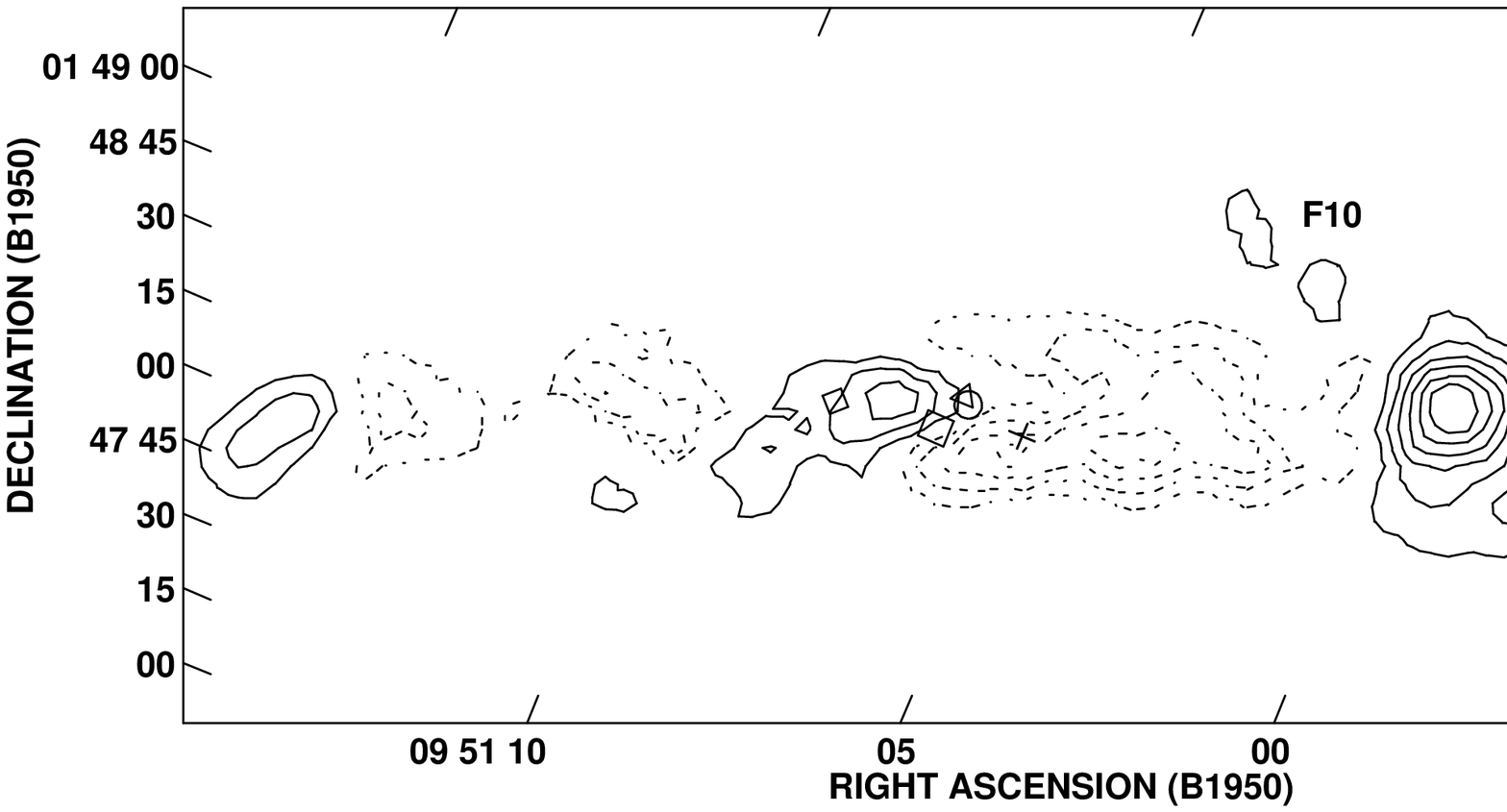}
\caption{Residual column density map obtained by subtracting
model from data. Solid contours represent emission above a smooth
distribution and dotted contours represent too much model emission.
Contour levels are at 36.1, 30.1, 24.1, 18.1, 12.0, 6.0, -6.0, -12.0,
-18.1 and -24.1$\times$10$^{20}$~cm$^{-2}$. Feature 10 is labelled as
F10 on the map. The symbols have the same meaning as in
Figure~\protect\ref{f4}.}
\label{f7}
\end{figure}

Figure~\ref{f7} shows the column density map of the residual cube for
the global model result. Negative (dotted) contours represent regions
where the amount of gas was over-estimated by the model and positive
(solid) contours represent regions of enhanced emission in the galaxy
above a smooth distribution. The peak column density in the residual
map is 30\% of the peak column density in the data
[Fig.~\ref{f3a}(a)], this shows that the model is able to reproduce
70\% of the underlying smooth distribution of the galaxy. There
remain, however, pixel-to-pixel variations between model and data due
to the unevenness of the distribution, as can be seen in
Fig.~\ref{f7}. For example, feature 10 appears as positive contours,
as do the two outer column density peaks [see also
Fig.~\ref{f3a}(a)]. The galaxy's nucleus derived from the best model
does not coincide with the central peak of the zeroth-moment map. As a
result, the model under-estimated the column density at the data peak.

\begin{figure}[ht]
\epsfxsize=3in\epsfbox[35 295 520 496]{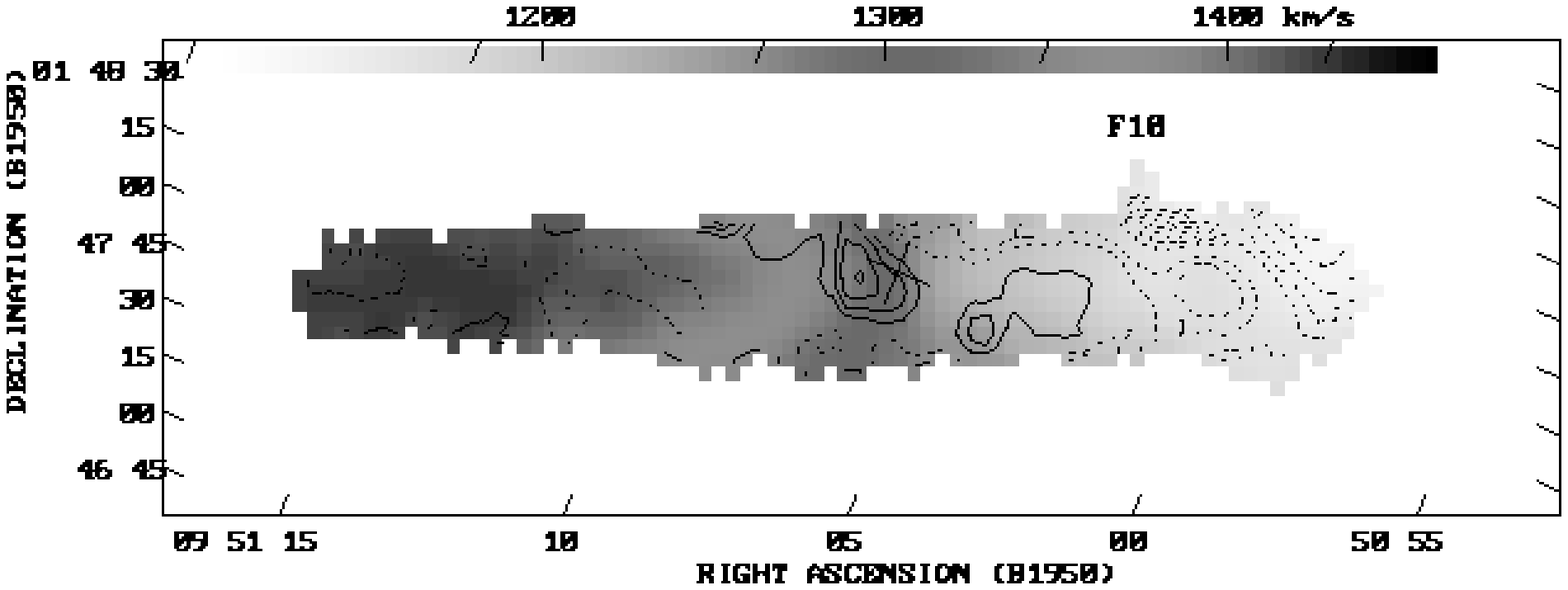}
\caption{Residual velocity field (contours) obtained by subtracting
the velocity field of the model from that of the data.  Contours are
at -55, -50, -45, -40, -35, -30, -25, -20, -10, -5, 5, 10, 15 and
20~\kms. It is superimposed on the grey scale image of the velocity
field of the data for comparison. Grey scale intensities are shown as
a band at the top. The cross marks the radio continuum centre of the
galaxy and F10 points to the location of feature 10.}
\label{f8}
\end{figure}

In terms of velocity, the model reproduced the velocity field of the
galaxy well for the most part. This is confirmed by subtracting the
first moment map of the model from that of the data
(Fig.~\ref{f8}). The rms of this residual velocity field is 16~\kms,
smaller than the velocity resolution of the data. There are regions in
the galaxy where the model's velocity cannot reproduce the data, a
good example is the region just under feature 10.  In this region, the
typical velocity excess is about -50~\kms\ (i.e., model over-estimated
the velocity) which represents the highest departure of the model from
the data. This is consistent with the observation made in
\S\ref{subsec:velocity}, which shows that the velocity field near
feature 10 is more blueshifted for its location than expected from
normal galactic rotation.

\section{The Remarkable Features of NGC~3044}
\label{sec:remarkable}

In the following, we discuss two of the most remarkable features of
NGC~3044, namely, the HI asymmetric distribution and the high-latitude
structures. These features are readily understood for an interacting
galaxy but remain puzzling in a seemingly isolated galaxy like
NGC~3044.

\subsection{Asymmetry}
\label{subsec:asymmetry}

\begin{figure*}[ht]
\hspace{1in}\epsfxsize=3in\epsfbox[79 281 533 710]{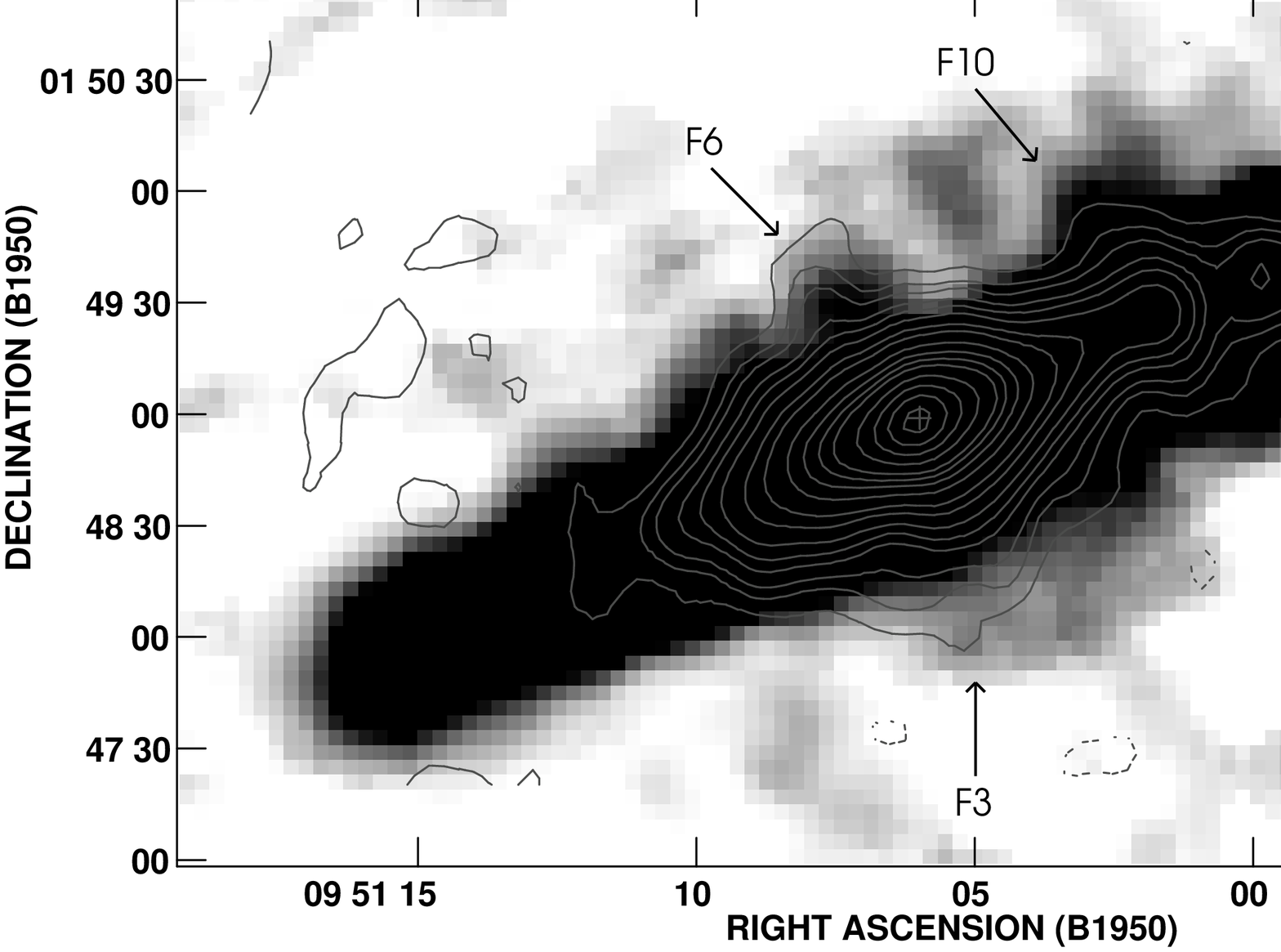}
\caption{20-cm C-array radio continuum contour map is superimposed on
the natural weighted HI column density map. The radio continuum data
has a beam size of $18\farcs 87\times 15\farcs 86$ and contours are at
-0.15, 0.15 (1$\sigma$), 0.30, 0.45, 0.75, 1.05, 1.50, 2.25, 3.00,
4.50, 6.00, 7.50, 9.00, 12.00, 15.00, 18.00, 21.00, and
22.50~mJy/beam. A cross marks the radio continuum centre of the
galaxy. Note that the HI disk has been ``burnt-out'' in order to show
the faint high-latitude features.}
\label{f9}
\end{figure*}

The asymmetric appearance of the optical disk has been described in
\S\ref{sec:intro} while the HI asymmetry is shown in the HI global
profile (Figure~5) and the rotation curve (Figure~\ref{f6}) and was
discussed in \S\ref{subsec:HIdistri}. In Figure~\ref{f9}, we compare
the HI distribution with the 20-cm radio continuum map of similar beam
size obtained from the completely independent data of Sorathia
(1994). There are remarkable correlations between the 2 maps, both
show extensions of the intensity contours towards the NW as well as
enhancements in intensity at the most north-westernly peak of the HI
column density map [$\alpha = 9^h51^m0\fs 0$, $\delta = 1\arcdeg
49\arcmin 35\farcs 8$ cf. Fig.~\ref{f3b}(b)].  All the above points to
a global asymmetry which manifested itself in various components of
this galaxy.

As indicated in \S~\ref{sec:intro}, there is no morphological evidence
for a disturbance in this galaxy (Solomon and Sage 1988).  What, then,
could have caused the asymmetric distribution in NGC~3044?  We note
here that HI asymmetries in field spirals are quite common. For
example, at least 50\% of 1400 field galaxies were classified as
asymmetrical by Richter and Sancisi (1994). However, the origin of the
asymmetry still eludes us. Zaritsky (1995) suggests that past mergers
may be the cause of the optical asymmetries in a sample of about 30
galaxies.  Could the asymmetry in NGC~3044 be similarly due to a past
merger, perhaps with a smaller companion?

Mergers between gas-rich disks and satellite galaxies (i.e., minor
mergers) have been studied extensively via simulations (e.g., Quinn
$\etal$, 1993, Mihos and Hernquist, 1994 and Hernquist and Mihos,
1995).  In general, the effect on the parent galaxy for assimilating a
satellite one tenth its own mass are: 1. massive gas in-flow to the
nucleus causing a brief ($\sim 10^8$ years) starburst phase;
2. heating of the stellar disk (i.e., increase in vertical scale
height by a factor of a few); 3. flaring and warping of the stellar
disk. In addition, depending on the satellite's initial density, its
core may or may not survive the tidal stripping to arrive at the
nucleus of the parent. Therefore, an obvious signature that a merger
has occurred is the existence of a double nucleus.  In addition,
Zaritsky (1995) finds a possible correlation between HI asymmetries
and star formation rates using the set of HI-asymmetric galaxies from
Rix and Zaritsky (1995).  The higher than normal star formation rates
are consistent with the starburst phase expected after a minor merger.

We do not yet have the data which could search for all of these
effects. Star formation rates can be investigated, though.  NGC~3044
is indeed classified as infrared-bright, but this is based on the IRAS
60$\mu m$ flux density (Soifer $\etal$ 1987). The infrared luminosity
calculated from the 60$\mu m$ and the 100$\mu m$ flux densities given
by Soifer $\etal$ (1989) is 4.8$\times10^9h^{-2}~L_{\sun}$ and the
massive star formation rate, following Condon (1992), is given by
$SFR(M\geq 5\solarmass) = 9.1\times10^{-11}\frac{L_{FIR}}{L_{\sun}} =
0.44h^{-2}~\solarmass /{yr}$. These values show that NGC~3044 is only
mildly starbursting at present (even if $h = 0.75$ is adopted, which
gives $SFR = 0.78~\solarmass /{yr}$).  That is, the massive star
formation rate in NGC~3044 is comparable to that of the Milky Way
(cf. 0.3 to 0.5 $\solarmass /{yr}$ for a supernova rate of 1 every 5 -
80 years) and lower than that of M~82 (2.2 $\solarmass /{yr}$) by a
factor of 5.  Thus if a minor merger has occurred in NGC~3044, it
likely occurred of order $\sim$ 10$^8$ years ago or more.  A study of
the CO content is underway to investigate the star formation
efficiency, determine whether the CO is also asymmetric and to search
for evidence of an increase in molecular gas density towards the
nucleus (Lee and Irwin, in preparation).  Further observations
(e.g. optical or IR searches for a double nucleus or a close
examination of the optical disk for flaring) would also be useful.

\subsection{High-Latitude Arcs and Extensions}
\label{subsec:arcs}
\subsubsection{Evidence of Expanding Shells}

The naturally weighted channel maps in Figure~\ref{f2a}(a) display
numerous low intensity, high-latitude arcs and extensions away from
the galactic disk.  These arcs and extensions in NGC~3044 resemble the
so-called ``Heiles Shells'' in our Galaxy. Table~\ref{tab:shells}
lists the positions, velocity ranges and the highest z-extent at the
1.5~$\sigma$ level of the more prominent extensions/holes seen in
Figure~\ref{f2a}(a). Only features which show up in more than one
consecutive channel are included. Some of these have very complex
appearances and are probably the results of blended features along the
same line-of-sight (e.g., feature 3). These features appear to be
distributed randomly along the disk of the galaxy.

Features 4 and 7 are clearly ``holes'' and are most obvious at
1339~\kms\ and 1297~\kms, respectively [see
Fig.~\ref{f2a}(a)]. Feature 6 straddles either side of the systemic
velocity (1287~\kms), hence we may be seeing both the receding and the
approaching caps of an expanding shell on the northeast side of the
galaxy (see below).  Feature 10 is the most massive (see
Table~\ref{tab:energetics}) extension in the list and spans 7 velocity
channels. At 1172 and 1152~\kms, feature 10 bends towards the east at
high latitude so that the feature looks like an arc. Feature 5 reaches
$8.4h^{-1}$~kpc above the midplane (most obvious at 1422~\kms\ where
it could be disconnected from the disk), making it the most extended
feature of all.

If any of these features are actually expanding shells or partial
shells, they should appear as rings or partial rings in the
position-velocity (P-V) slices across the centres of the
features. Figure~10(a) and (b) show the zeroth-moment maps of features
4, 7, 10 and 12 and their associated P-V diagrams (natural weighting
data). Each slice is an average of 2 pixels (8$\arcsec$) parallel to
the major axis. In each P-V slice shown in Figs.~10, we find expansion
signatures at the position and velocity corresponding to these
features. In all cases, the shells are not complete but are ``open''
at the sides toward lower density, both in P-V and in RA-DEC
space. This is consistent with the Chimney models for a blow out case
(see below).  It could also indicate that the shell has more readily
fragmented in low density regions.

\begin{figure*}
\epsfxsize=5in\epsfbox[80 330 475 620]{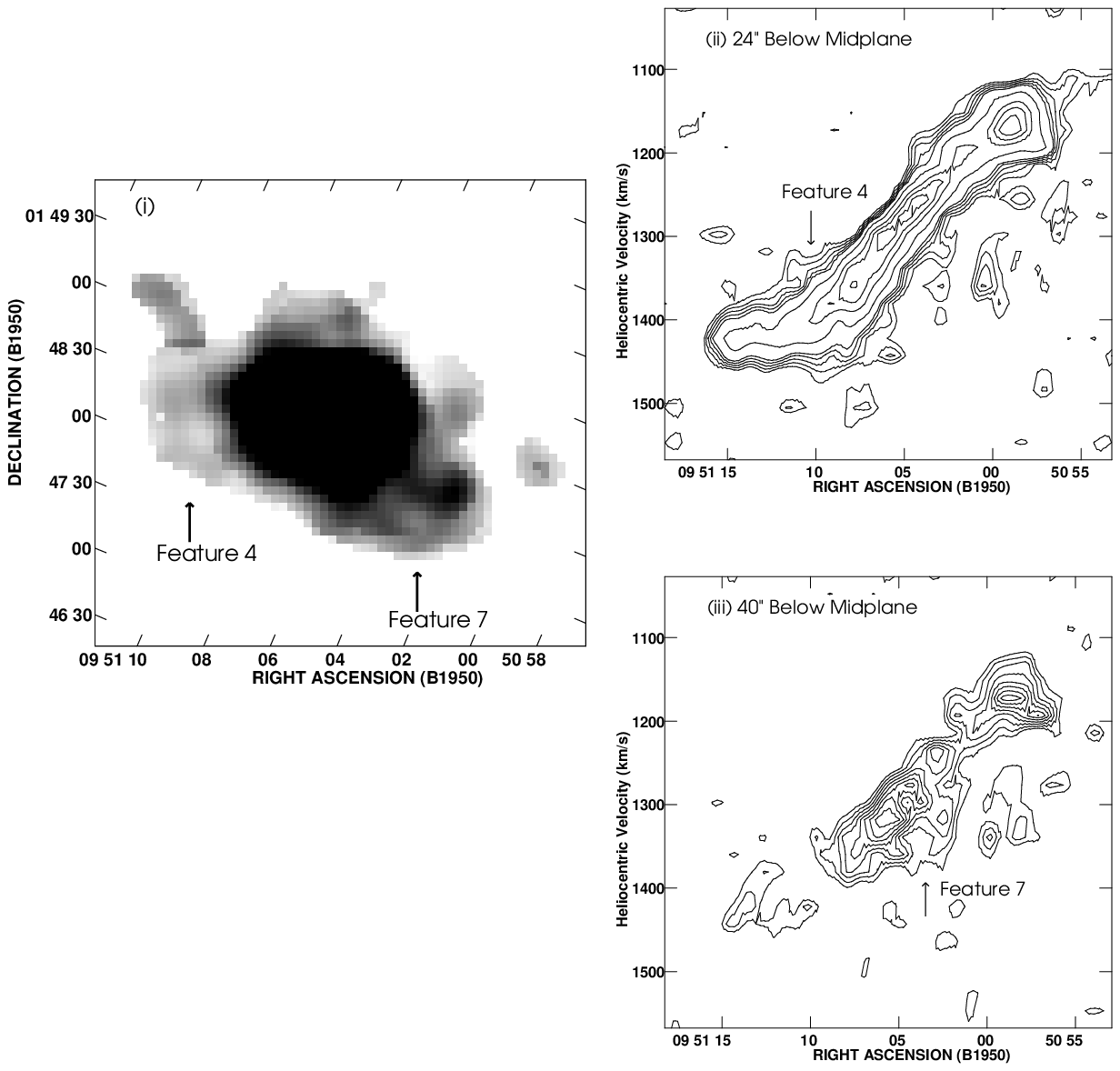}
\caption{(a) (i)Zeroth-moment of the velocity channels showing
features 4 and 7. Smoothing is done as in
Figs.~\protect\ref{f3a}. (ii) and (iii) show position-velocity (P-V)
slices parallel to the major axis across feature 4 and 7,
respectively. Each slice is an average of 2 pixels
(8$\arcsec$). Contours are at 0.64 (1$\sigma$), 1, 1.3, 1.6, 1.9, 3.2,
5.1, 6.4, 7.7~mJy/beam. The height above and below the midplane is
shown in each P-V panel.}
\label{f10a}
\end{figure*}

\addtocounter{figure}{-1}

\begin{figure*}
\epsfxsize=4.5in\epsfbox[100 230 431 535]{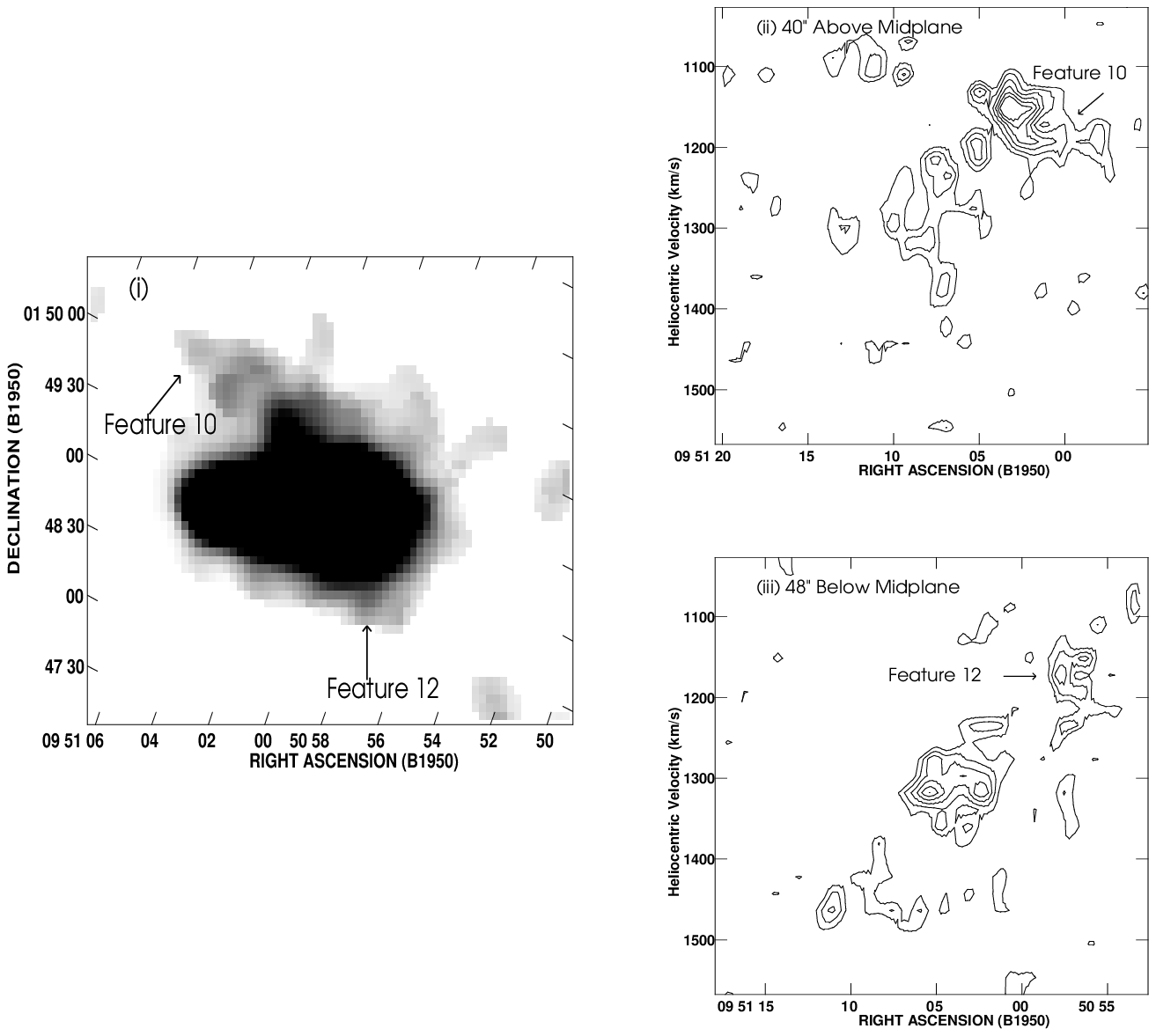}
\caption{(b) as in (a) but for feature 10 and 12.}
\label{f10b}
\end{figure*}

All the features in Table~\ref{tab:shells} reach a projected height of
at least 5 times the HI scale-height derived from our model (see
Table~\ref{tab:model}. It is clear that, whatever the mechanism which
created them, it must be energetic enough that the features are able
to break through the moderately thick HI disk to reach the halo (this
is termed {\it blowout} in \cite{hei90}). For a blowout case, mass and
momentum are injected into the halo.  The total mass of all features
measured is $2\times10^8h^{-2}~\solarmass$ (a lower limit, see next
section) or 7\% of the total HI mass of the galaxy.  If we assume that
10\% of this mass eventually reaches the halo, and the typical
lifetime for these features is about $3\times10^7h^{-1}$~years (see
$\tau_{sh}$ in Table~\ref{tab:energetics} and the next section), then
a lower limit to the mass injection rate for cold neutral gas,
$\dot{\it M}$, is $0.7h^{-1}~\solarmass/yr$.  This is in rough
agreement with numerical calculations utilizing the ISM parameters of
the Galaxy (\cite{hei90}).

\subsubsection{Supershell Parameters}

Table~\ref{tab:energetics} lists the masses ({\it M$_{sh}$}, column
2), expansion velocities ({\it V$_{sh}$}, column 3), radii ({\it
R$_{sh}$}, column 4), projected galactocentric locations ({\it
D$_{sh}$}, column 5), ambient densities in the mid-plane at {\it
D$_{sh}$} ({\it n$_0$}, column 6) , and kinetic energies ({\it E$_k =
\case{1}{2}M_{sh}V_{sh}^2$}, column 7) of the expanding shells. The
masses are obtained by summing the fluxes associated with each
features in the velocity channels listed in
Table~\ref{tab:shells}. These masses are lower limits since part of
the shells are most certainly embedded within the HI disk and are not
included. For holes 4 and 7, the masses are not listed as the
thickness of the shell cannot be determined with confidence. {\it
V$_{sh}$} is taken to be half the total velocity range which again may
underestimate the actual range. Radii of the shells are measured from
the 1.5$\sigma$ contours of the P-V diagrams. {\it D$_{sh}$} is an
average of the measurements made from the channel maps and the P-V
diagrams. We use the modelled density profile in the plane of the
galaxy (see column 2 of Table~\ref{tab:model}) to estimate the ambient
density at the galactocentric distance of a given shell.

\begin{deluxetable}{lccccccc}
\tablewidth{0pt}
\tablecaption{Energy Requirements of Expanding Supershells 
\label{tab:energetics}}
\tablehead{
\colhead{Feature} & \colhead{\it M$_{sh}$} & \colhead{\it V$_{sh}$} &
\colhead{\it R$_{sh}$} & \colhead{\it D$_{sh}$} & 
\colhead{\it n$_0$} & \colhead{\it E$_k$} & \colhead{\it E$_E$} \nl
\colhead{} & \colhead{$h^{-2}~\solarmass$} & \colhead{\kms} &
\colhead{$h^{-1}~kpc$} & \colhead{$h^{-1}~kpc$} & 
\colhead{$h~\cubiccm$} & \colhead{$h^{-2}$~ergs} & \colhead{$h^{-2}$~ergs} \nl
\colhead{(1)} & \colhead{(2)} & \colhead{(3)} & \colhead{(4)} & \colhead{(5)} &
\colhead{(6)} & \colhead{(7)} & \colhead{(8)}}
\startdata
\nl
4 & \nodata & 31.2 & 1.2 & 6.7 & 0.18 & \nodata & 4.1$\times$10$^{54}$ \nl
7 & \nodata & 31.2 & 1.7 & 0.0 & 0.01 & \nodata & 4.1$\times$10$^{53}$ \nl
10 & 5.5$\times$10$^7$ & 72.8 & 2.0 & 6.3 & 0.21 & 
	2.9$\times$10$^{54}$ & 7.4$\times$10$^{55}$ \nl
12 & 1.6$\times$10$^7$ & 41.6 & 1.5 & 10.1 & 0.04 & 
	2.7$\times$10$^{53}$ & 1.9$\times$10$^{54}$ \nl
\nl

\enddata
\end{deluxetable}

The kinematical age of a shell, assuming constant expansion velocity,
is $\tau_{sh}=R_{sh}/V_{sh}$. For the four shells listed in
Table~\ref{tab:energetics}, we expect $V_{sh}$ to be underestimated
(see above) and therefore $\tau_{sh}$ may be overestimated.  The
resulting kinematical ages range from 24 million years for shell 10,
to 52 million years for shell 7.  These ages are comparable to the
expected lifetime of an OB association.

For an expanding shell which is formed from
a one-time energy injection, such as from supernovae
and is now in the radiating phase of its
evolution, numerical analysis by Chevalier (1974) shows that the
energy injected is given by
\begin{displaymath}
E_E = 5.3\times 10^{43}n_0^{1.12}R_{sh}^{3.12}V_{sh}^{1.4}
\end{displaymath}
where the variables have the same meaning as before ($R_{sh}$ is in pc
and $V_{sh}$ in \kms).  Values of {\it E$_E$} for the expanding shells
in NGC~3044 are listed in column 8 of Table~\ref{tab:energetics}.  As
Table~\ref{tab:energetics} shows, since the energies are all upwards
of 10$^{53}$~ergs, it is evident that all four shells are
``supershells'', as defined by Heiles (1979, i.e. with energies $>$ 3
$\times$ 10$^{52}$ ergs).

Since the input energy requirements for supershells are significant,
it is worth considering the errors on $E_E$.  As indicated previously,
the measured $V_{sh}$ is a lower limit since complete shells are not
observed.  Increasing $V_{sh}$ by a factor of 2, for example, would
increase $E_E$ by a factor of 2.6.  We have also minimized $R_{sh}$
through our choice of Hubble constant (see dependence on $h$ in
Table~\ref{tab:energetics}).  Thus a change to $h = 0.75$ increases
$E_E$ by a factor of 1.8.  If the features are actually located at
larger galactocentric distances than the projected radius, $D_{sh}$,
then the ambient densities, $n_0$, would be lower than the listed
values.  In the extreme case of a shell actually occurring at the
outer radius of the HI disk ({\it{R}} = $13.4h^{-1}~kpc$), then $n_0 =
0.01h~\cubiccm$, decreasing $E_E$ by a factor of 30 for feature 10,
which has the highest $n_0$ in Table~\ref{tab:energetics}.  However,
note that for Feature 10, $E_k$ = 3 $\times$ 10$^{54}$ ergs.  This
value gives only the kinetic energy of expansion of the feature and
thus the input energy must be much higher (i.e. by at least a factor
of 10 for an efficiency of 10\%).  Therefore, $E_E$ for this feature
is likely to be within a factor of 2 of the value listed in
Table~\ref{tab:energetics}.  This then implies that Feature 10 is at a
position much closer to its projected galactocentric distance than
near the edge of the disk.  Given the irregular density distribution
in the galaxy (\S~\ref{subsec:residual}, Fig. 7), there are also
possible variations in $n_0$ for individual shells.  For example,
Feature 7, with a projected center at the center of the galaxy, is the
most extreme example since it could have an energy more than an order
of magnitude larger than the table~\ref{tab:energetics} value if $n_0$
is determined by the fits to the two sides of the galaxy separately,
rather than to the galaxy as a whole (Table~\ref{tab:model}).
Overall, however, the irregularities in density tend to be within
$\sim$ 30\% of the modelled density, introducing similar sized errors
into $E_E$.  Most of the assumptions we have made effectively minimize
the values of $E_E$ in Table~\ref{tab:energetics}.  We conclude that
the energy estimates, under the assumption of instantaneous input, are
order of magnitude values.

 If supernova explosions are the source of the energy, then between
$\approx 400$ to $70,000$ supernovae are required to produce the
shells in Table~\ref{tab:energetics}. Heiles (1979) found energies an
order of magnitude lower for Galactic shells while shells in NGC~3079
(\cite{irw90}) and NGC~4631 (\cite{ran93}) require comparable energies
($10^{54}$ - $10^{55}$~erg) when scaled to $h = 1$. Note that NGC~3044
probably has many lower energy shells like those in the Galaxy and the
fact that we detect only the supershells is merely a selection
effect. Such a high energy requirement for the supershells is
difficult to reconcile with input energies typical of Galactic OB
associations which contain only a few tens of stars of spectral types
B0 and earlier.  H$\alpha$ luminosities of bright HII regions in some
external late-type galaxies do suggest the existence of superclusters
which contain thousands of supernovae (Heiles 1990) and therefore
H$\alpha$ observations of NGC~3044 may be helpful in this regard.

There is certainly some dependence of the required input energy on
adopted model.  For example, if a slower, continuous energy injection
is assumed over a typical cluster ($\approx$ shell) age (several
$\times$10$^7$ years), then we find energies which agree to within a
factor of 4 with the tabulated $E_E$ for Features 4 and 10, but are at
least an order of magnitude lower for the low density Features 7 and
12 (cf. Vader and Chaboyer 1995).  Also, since the supershells have
achieved blowout, an understanding of these features would also
benefit from numerical hydrodynamical modelling for such conditions
(cf. Mac Low, McCray, \& Norman 1989).  Nevertheless, one cannot
escape the fact that some of the observed high latitude features in
NGC~3044 are supershells which require large input energies to
produce.  The kinetic energies of expansion, $E_k$, which are
model-independent and essentially assume 100\% input efficiency, are
of order 10$^{53}$ - 10$^{54}$ ergs and these kinetic energies are
underestimated for reasons outlined above.  For a more realistic
situation in which the efficiency is 10\% or less, the input energies
must be at least an order of magnitude larger.

\subsubsection{Origin of the Supershells}
\label{subsubsec:shellorigin}

We have so far implicitly assumed that the shells are formed
internally through the collective effects of clustered
supernovae. Since the energy requirement are very high, however, the
leading alternative scenario, i.e. that the supershells are produced
from impacting external clouds, should also be considered.

Impacting cloud models are an attractive way to explain large,
energetic supershells because the resulting energies are a function of
the infalling mass for which there are no hard limits in the case of
galaxy-galaxy interactions.  If we assume that the supershells in
NGC~3044 are due to impacting clouds, however, we must consider where
such clouds would originate.  NGC~3044 has no nearby companion, nor is
there strong evidence (apart from the asymmetry) for a previous
interaction.  Moreover, a previous interaction, if it occurred, more
likely occurred over timescales of order 10$^8$ yr (a typical
interaction timescale, see also \S5.1).  However, we find kinematical
ages for the shells of a few $\times$ 10$^7$ yrs.  Therefore, if the
shells are produced by infalling clouds, the situation is more likely
one in which high velocity clouds (HVCs) exist around the galaxy and
are continuously ``raining down''.  This also suggests that there
should be evidence for such clouds around the galaxy now.

One of the two supershells found in NGC~4631 (\cite{ran93}) was
modelled by Rand and Stone (1996) as a HVC impact structure using a
3-D hydrodynamical simulation.  They found the most likely HVC that
formed the supershell has a radius of 500~pc and an HI mass of
$1.2\times10^7~\solarmass$. Since NGC~3044 and NGC~4631 have
remarkably similar kinematics, global HI distribution and shell
parameters (see \cite{ran94}), we can assume that infalling clouds of
similar size and mass are required to form the supershells in
NGC~3044.  Converting these parameters to a column density, we find
(within one beam) a value of $N_{HI} = 5.2\times10^{20}~cm^{-2}$,
which corresponds to the fifth contour in Figure~3(b).  There is
clearly no evidence at the present time for such massive clouds in the
vicinity of NGC~3044.  Thus, the impacting cloud model may be
reasonable for interacting galaxies like NGC~4631, but is much less
attractive for an isolated system like NGC~3044.

From Figure~9, there appears to be a correlation between high latitude
radio continuum and high latitude HI features in the case of features
3, 6, and 10. Indeed there appears to be an HI feature wherever a
radio continuum feature exists. However, the converse is not true, as
there are a number of HI features which do not (at the sensitivity
limit of the observations) have corresponding radio continuum
features.  The observations are not inconsistent, at least
qualitatively, with the Chimney model (\cite{nor89}).  In this model,
the HI features are walls of ambient material swept up by spatially
correlated supernova explosions. Relativistic electrons then funnel
through these ``Chimneys'' to reach the halo.  It is difficult to
tell, at this resolution and with the blended features, whether the
radio continuum emission is interior to a specific chimney, or whether
the emission is coincident with the walls of a chimney.

As the above analysis shows, the formation of supershells in NGC~3044
is far from clear. Both supernovae models and the cloud-galaxy
collision model have difficulties.  However, given the timescales for
shell formation, the absence of companion galaxies near NGC~3044, and
the lack of evidence for surrounding massive clouds, and the apparent
correlation between high latitude HI and radio continuum features, we
favour an internal origin for the supershells.  Several recent studies
have shown, for example, that when magnetic fields are included (not
just as a secondary effect) that large shells and blow-out can occur
with more modest energy requirements (e.g. Kamaya $\etal$ 1996; Frei
$\etal$ 1997).  Therefore, more sophisticated models including effects
like this may reconcile the energy requirements of supershells with
fewer numbers of supernovae.

\section{Summary}
\label{sec:summary}

What has been presented in the preceding sections represents the first
comprehensive HI study of the {\it isolated}, edge-on, infrared-bright
galaxy NGC~3044.  The IR brightness is partly due to proximity, since
the massive star formation rate is a factor of 5 lower than that of
M~82 (\S 5.1).

The HI distribution in the galaxy was modelled using all available
data points, i.e. HI spectral lines were modelled over the entire
galaxy in order to determine the best global density and velocity
distributions in the galaxy as well as the best geometrical
parameters.  The resulting quantities are listed in
Table~\ref{tab:model}.  A Gaussian ring density distribution in the
plane was found to be the best fit overall.  An important result
derived from the model is that the galaxy has a moderately thick HI
disk with a Gaussian vertical scale height of $420h^{-1}~pc$.  This is
consistent with previous observations which have also revealed a thick
radio continuum disk.

NGC~3044 shows a very asymmetric distribution optically, in HI, and in
the radio continuum, with excellent agreement between the radio
continuum and HI column density distributions (see
\S\ref{subsec:asymmetry}).  In HI, the asymmetry is particularly
apparent from the global profile, which shows a lopsided double-horned
structure.  Otherwise, however, derived global parameters tell the
story of a normal SBc galaxy.  The column density map confirms the
asymmetric distribution of the HI in the galaxy. This is clear from
the 27$\arcsec$ offset of the HI peak to the east of the radio
continuum peak.  The major-axis rotation curve also shows the
asymmetry in that the approaching side of the galaxy appears to be
``truncated'' before reaching terminal velocity while the receding
side stays flat for about $7h^{-1}~kpc$ after reaching terminal
velocity.  While the galaxy is asymmetric in HI, it also shows an
extremely straight disk (i.e., no obvious disk warp).

It is not clear what could have caused the asymmetry in this
apparently isolated galaxy.  It is likely, though, that the mechanism
would have to act on the galaxy after its formation, since it is
difficult to explain how the galaxy would form asymmetrically in the
first place.  It is noted here that asymmetries in isolated galaxies
are not uncommon, which indicates that a common mechanism such as past
minor mergers may be responsible for it.  If this is the case for
NGC~3044, the merger likely took place over 10$^8$ yrs ago.

The velocity channel maps reveal a host of high-latitude HI
extensions.  We have catalogued a total of 12 such features that
appear in at least two consecutive velocity channels. These features
are distributed uniformly across the disk of the galaxy and above and
below the plane. We derive a lower limit for mass flow to the halo of
$\dot{\it M}$, = $0.7h^{-1}~\solarmass/yr$.  There is also some
correlation between high-latitude HI features and high-latitude radio
continuum features ($\S$ 5.2.3).

Four of the high latitude features exhibit the signature of an
expanding shell and have radii of order a few kpc.  Based on the
assumption that the shells were formed by supernovae energy input, the
required input energies are in the range 4 $\times$ 10$^{53}$ to 7
$\times$ 10$^{54}$ $h^{-2}$ ergs.  These values vary somewhat
depending on the adopted model, but it is clear, even from the kinetic
energies of expansion of the individual supershells, that large
numbers (e.g. up to tens of thousands) of correlated supernovae are
required to form them.  This means that NGC~3044 would have to harbour
a number of such superclusters (such as the cluster R136 in the 30
Doradus region in the LMC), something which is not supported by the
current massive star formation rate.

The alternative model, that of impacting external clouds, is even less
attractive as a formation mechanism for the supershells because of the
absence of companions, the fact that the supershells are young (of
order the age of a typical OB association) and because HI clouds which
are massive enough to produce these shells would have been easily
detected in our data.  We suggest that the supershells are indeed
formed internally, but that some additional energy-boost (e.g. through
magnetic fields) is needed to explain their large energies.

\acknowledgments The authors wish to thank the staff at the VLA for
obtaining the HI data and Kathy Perrett for coordinating the remote
observation. Thanks also go to Denise Giguere and Jayanne English for
many useful discussions on various data reduction techniques. We have
also made used of the NASA/IPAC Extragalactic Database (NED). This
work is supported by the National Research Council of Canada.


\end{document}